\documentclass[12pt]{article}
\pdfoutput=1
\usepackage{jheppub_X}

\usepackage[T1]{fontenc}
\usepackage[utf8]{inputenc}
\usepackage{dsfont}
\usepackage{cleveref}
\crefname{table}{Table}{Tables}
\crefname{equation}{Eq.}{Eqs.}
\crefname{appendix}{App.}{Apps.}
\crefname{section}{Sec.}{Secs.}
\crefname{figure}{Fig.}{Figs.}
\usepackage{amsmath}
\usepackage{setspace}

\usepackage{bm}
\usepackage{xfrac}
\usepackage{pbox}
\usepackage{comment}
\usepackage{slashed}
\usepackage{physics}
\usepackage{enumitem}
\usepackage{mathrsfs}

\usepackage[dvipsnames,table]{xcolor}
\usepackage[most]{tcolorbox}
\definecolor{light-gray}{gray}{0.9}
\usepackage{bbold}

\usepackage[normalem]{ulem}

\makeatletter
\g@addto@macro\bfseries{\boldmath}
\makeatother

\newcommand{\ord}{{\mathcal{O}}}
\newcommand{\Lag}{{\mathcal{L}}}
\newcommand{\Dcal}{{\mathcal{D}}}
\newcommand{\Amp}{{\mathcal{A}}}
\newcommand{\Mcal}{{\mathcal{M}}}
\newcommand{\Chris}{{G}}
\newcommand{\tphi}{{\tilde\phi}}
\newcommand{\tpsi}{{\tilde\psi}}
\newcommand{\teta}{\tilde{\eta}}
\newcommand{\tGamma}{{\widetilde\Gamma}}
\newcommand{\tChris}{{\widetilde\Chris}}
\newcommand{\tMcal}{{\widetilde{\mathcal{M}}}}
\newcommand{\tAmp}{{\widetilde{\mathcal{A}}}}

\newcommand{\tJ}{{\tilde{J}}}
\newcommand{\pon}{\bar{p}}
\newcommand{\qon}{\bar{q}}

\def\eg{\textit{e.g.}}
\def\ie{\textit{i.e.}}
\def\etc{\textit{etc. }}
\def\cf{\textit{c.f.}}

\newcommand{\tZ}{\widetilde{Z}}
\newcommand{\expv}[1]{\left\langle #1 \right\rangle}
\newcommand{\tD}{\widetilde{D}}
\newcommand{\tG}{\widetilde{G}}
\newcommand{\tcM}{\widetilde{\mathcal{M}}}
\newcommand{\cM}{\mathcal{M}}
\newcommand{\tnabla}{\widetilde{\nabla}}

\newcommand{\dotrad}{0.43}
\newcommand{\dotradM}{0.4}

\tikzset{vertexstyle/.style={circle,draw,thick,inner sep=2pt}}
\tikzset{dotstyle/.style={thick,dotted}}
\tikzset{propstyle/.style={ultra thick}}

\definecolor{colorTC}{rgb}{.2,.7,.2}

\newcommand{\s}{\hspace{0.8pt}}

\preprint{CERN-TH-2023-233}

\title{\Large On Amplitudes and Field Redefinitions}

\author[a,b,c]{Timothy Cohen,}
\author[d]{Xiaochuan Lu,}
\author[e]{and Dave Sutherland\s}

\affiliation[a]{\fontsize{10}{10}\selectfont Theoretical Physics Department, CERN, 1211 Geneva, CH}
\affiliation[b]{\fontsize{10}{10}\selectfont Theoretical Particle Physics Laboratory, EPFL, 1015 Lausanne, CH}
\affiliation[c]{\fontsize{10}{10}\selectfont Institute for Fundamental Science, University of Oregon, Eugene, Oregon 97403, USA}
\affiliation[d]{\fontsize{10}{10}\selectfont Department of Physics, University of California, San Diego, La Jolla, CA 92093, USA}
\affiliation[e]{\fontsize{10}{10}\selectfont School of Physics and Astronomy, University of Glasgow, Glasgow G12 8QQ, UK}

\emailAdd{tim.cohen@cern.ch}
\emailAdd{xil224@ucsd.edu}
\emailAdd{david.w.sutherland@glasgow.ac.uk}

\abstract{
We derive an off-shell recursion relation for correlators that holds at all loop orders. This allows us to prove how generalized amplitudes transform under generic field redefinitions, starting from an assumed behavior of the one-particle-irreducible effective action. The form of the recursion relation resembles the operation of raising the rank of a tensor by acting with a covariant derivative. This inspires a geometric interpretation, whose features and flaws we investigate.
}

\begin{document}
\maketitle
\flushbottom
\setcounter{page}{2}
\newpage

\begin{spacing}{1.1}
\parskip=0ex

%%%%%%%%%%%%%%%%%%%%%%%%%%%%%%%%%%%%%%%%%%%%%%%%%%%%%%%%%%%%%%%%%%%%%%%%%%%%%%%%
\section{Introduction}
\label{sec:Introduction}
%%%%%%%%%%%%%%%%%%%%%%%%%%%%%%%%%%%%%%%%%%%%%%%%%%%%%%%%%%%%%%%%%%%%%%%%%%%%%%%%

It is well understood that defining a theory in terms of fields introduces a tremendous redundancy. In particular, one of the most fundamental quantities that can be computed from a field theory are the $S$-matrix elements or amplitudes. Amplitudes are known to be invariant under field redefinitions of the form \cite{Chisholm:1961tha, Kamefuchi:1961sb, Arzt:1993gz, Manohar:2018aog}
\begin{align}
\phi(x) \quad\longrightarrow\quad \phi(x) + f \big( \phi(x),\, \partial_\mu \phi(x),\, \partial_\mu \partial_\nu \phi(x),\, \cdots \big) \,,
\label{eq:generalFieldRedef}
\end{align}
where $f$ is an arbitrary polynomial function of the field(s) and its derivatives evaluated at the spacetime point $x$. This field redefinition invariance plays a minor role for ``renormalizable'' theories (with the important exception of gauge theory). However, this redundancy becomes a significant source of technical complexity when one studies ``non-renormalizable'' Effective Field Theories (EFTs) that include irrelevant operators. In the case of EFTs, the ability to perform field redefinitions, often expressed as an iterative equation of motion redundancy (along with the application of integration by parts) implies that the space of allowed operators is highly redundant, so that Lagrangians which appear different actually describe the same underlying scattering physics.

In this paper, we build upon and explore the results in Ref.~\cite{Cohen:2022uuw} to provide a new perspective on the notion of field redefinition invariance of amplitudes.  In particular, we prove a ``transformation lemma'' for an off-shell generalization of the amplitudes. We then apply this result to show that the on-shell amplitudes are invariant under field redefinitions (up to one-loop order). This new approach follows as a direct consequence of a new off-shell recursion relation that we prove in this paper.

The study of field redefinitions and EFTs has undergone something of a renaissance in recent years. The determination of the size of the operator basis using the Hilbert series has been developed and applied to many examples~\cite{Lehman:2015via, Henning:2015daa, Lehman:2015coa, Henning:2015alf, Kobach:2017xkw, Henning:2017fpj, Kobach:2018pie, Ruhdorfer:2019qmk, Marinissen:2020jmb, Graf:2020yxt, Graf:2022rco, Sun:2022aag, Kondo:2022wcw, Delgado:2022bho, Sun:2022snw, Bijnens:2022zqo, Delgado:2023ivp, Grojean:2023tsd}. This is closely related to the approach of constructing EFT amplitudes directly~\cite{Cohen:2010mi, Cheung:2015aba, Azatov:2016sqh,  Arkani-Hamed:2017jhn, Shadmi:2018xan, Christensen:2018zcq, Durieux:2019eor, Durieux:2019siw, Bern:2019wie, Christensen:2019mch, Ma:2019gtx, Aoude:2019tzn, Bachu:2019ehv, Henning:2019enq, Bern:2020ikv, Durieux:2020gip, EliasMiro:2020tdv, Baratella:2020lzz, Falkowski:2020fsu, Jiang:2020mhe, Jin:2020pwh, Nagai:2021gmo, Dong:2021yak, AccettulliHuber:2021uoa, DeAngelis:2022qco, Chang:2022crb, Dong:2022mcv, Balkin:2021dko, Low:2022iim, Liu:2023jbq, Bradshaw:2023wco, Arzate:2023swz}, which again avoids the issues of operator redundancies. In both approaches, the full set of field redefinitions included in \cref{eq:generalFieldRedef} are accommodated.

Another fruitful approach is to work with the Lagrangian directly, but to express it in terms of geometric objects defined on a Riemannian field space manifold~\cite{Honerkamp:1971sh, Tataru:1975ys, Alvarez-Gaume:1981exa, Alvarez-Gaume:1981exv, Vilkovisky:1984st, DeWitt:1984sjp, Gaillard:1985uh, DeWitt:1985sg}. In this case, the key insight is to identify that field redefinitions without derivatives are equivalent to coordinate changes on the field space manifold. One can then express amplitudes directly in terms of well known geometric quantities built out of the Riemannian metric. This makes the invariance of amplitudes under the restricted set of field redefinitions completely manifest. This approach has seen recent applications to the scalar sector of the Standard Model~\cite{Alonso:2015fsp, Alonso:2016btr, Alonso:2016oah, Nagai:2019tgi, Helset:2020yio, Cohen:2020xca, Cohen:2021ucp, Alonso:2021rac, Banta:2021dek, Talbert:2022unj, Alonso:2023jsi}, and has also led to new insights into the properties of amplitudes for both scalars and particles of higher spins~\cite{Finn:2019aip, Finn:2020nvn, Cheung:2021yog, Alonso:2022ffe, Helset:2022tlf, Helset:2022pde, Pilaftsis:2022las, Assi:2023zid, Jenkins:2023rtg, Jenkins:2023bls, Gattus:2023gep, Alonso:2023upf}.

However, the geometric picture that can accommodate the full set of field redefinitions has remained elusive \cite{Cohen:2022uuw, Cheung:2022vnd, Craig:2023wni, Craig:2023hhp, Alminawi:2023qtf}. It is this search for a generalized notion of geometry that has prompted us to revisit the field redefinition properties of amplitudes.  In particular, we will show that our new perspective has a natural geometry-like interpretation, that we call ``functional geometry.'' We are able to find hints that functional geometry exists and has the desired features to be associated with a generalized manifold. However, we also show that it fails to fully generalize field space geometry in a number of important ways. Nevertheless, we are optimistic that the ideas presented here represent genuine progress towards what will eventually be the discovery of a new notion of geometry that accommodates the full set of allowed field redefinitions.

The rest of this paper is organized as the following. In \cref{sec:Recursion}, we review the well-known path integral formalism of QFTs and use it to derive a recursion relation for the off-shell amplitudes that holds at all loop orders. We then make use of it to prove a transformation lemma in \cref{sec:Invariance}, which is the main result of this paper. We demonstrate that up to one-loop level, this lemma applies when a general field redefinition that accommodates derivatives is taken, which then immediately implies the invariance of the on-shell amplitudes. In \cref{sec:Geometry}, we present an attempt to introduce a geometric interpretation, motivated by the tensor-like structure of the recursion relation derived in \cref{sec:Recursion}. In particular, we discuss the successes and failures of this interpretation, and comment on its relation with the well-established field space geometry picture. Conclusions and future directions are given in \cref{sec:Conclusions}.

%%%%%%%%%%%%%%%%%%%%%%%%%%%%%%%%%%%%%%%%%%%%%%%%%%%%%%%%%%%%%%%%%%%%%%%%%%%%%%%%
\section{Off-shell Recursion for Amplitudes}
\label{sec:Recursion}
%%%%%%%%%%%%%%%%%%%%%%%%%%%%%%%%%%%%%%%%%%%%%%%%%%%%%%%%%%%%%%%%%%%%%%%%%%%%%%%%

We begin with a brief review of the formalism for computing correlation functions from the path integral  (\cref{subsec:Correlation}). The partition function $Z[J]$ spans a set of theories that are parameterized by difference choices of the source fields $J(x)$, and the original theory corresponds to taking $J(x)=0$, \ie, the ``zero source condition.'' We then review the LSZ formalism for projecting amplitudes from the correlation functions (\cref{subsec:Amplitudes}). The LSZ formula provides a general definition of ``amplitudes'' which allow for the external states to be off-shell;  the limit where the external states are on-shell defines the ``$S$-matrix elements.'' Although these first two subsections contain material typically covered in QFT textbooks, our purpose here is to express these well-known results in a notation that is convenient for deriving a recursion relation for off-shell amplitudes that holds at all loop orders \cite{Cohen:2022uuw} (\cref{subsec:Recursion}).

\clearpage

\subsection{Correlation Functions From the Path Integral}
\label{subsec:Correlation}

Given a scalar field $\eta(x)$, whose action is given by $S[\eta]$, one can define the partition function as a path integral
\begin{equation}
Z[J] \equiv e^{iW[J]} \equiv \int \Dcal\s\eta\, e^{iS[\eta] +\s  i\int \dd^4x\, J(x)\s \eta(x)} \,,
\label{eqn:ZWdef}
\end{equation}
and we have defined $iW[J]\equiv \log Z[J]$ as usual. The partition function $Z[J]$ is a generating functional of the (time-ordered) $J$-dependent correction functions
\begin{equation}
\langle \eta^{x_1} \cdots \eta^{x_n} \rangle_{J} \equiv \frac{\int \Dcal\s\eta\, e^{iS[\eta] + iJ_x\eta^x}\, \eta(x_1) \cdots \eta(x_n)}{\int \Dcal\s\eta\, e^{iS[\eta] + iJ_x\eta^x}} = \frac{1}{Z[J]} (-i)^n \frac{\delta^n Z}{\delta J_{x_1} \cdots \delta J_{x_n}} \,,
\label{eqn:CorrJ}
\end{equation}
where we have introduced the concise notation\footnote{We will use both notations in what follows based on convenience.}
\begin{subequations}\label{eqn:xConcise}
\begin{align}
\eta(x) &= \eta^x \,, \\[5pt]
J(x) &= J_x \,,
\end{align}
\end{subequations}
so that an integral over spacetime is represented as a sum over a dummy index
\begin{equation}
\int \dd^4x\, J(x)\, \eta(x) = J_x \eta^x \,.
\label{eqn:xSum}
\end{equation}

It is well known that the path integral formalism and the use of generating functionals being reviewed in this section generalizes to an arbitrary set of bosonic and fermionic fields \cite{Neufeld:1998js, Weinberg:1996kr, Schwartz:2014sze}. When dealing with fermionic fields, one needs to keep track of the signs carefully. In the case of a general field, the index $x$ in \cref{eqn:xConcise} is understood to collectively label the spacetime position, the spin indices, as well as any of its internal flavor indices, all of which are summed over when the dummy index $x$ is contracted.

\subsubsection*{Source Dependence}

The $J$-dependent correlation functions $\langle \eta^{x_1} \cdots \eta^{x_n} \rangle_{J}$ can be viewed as the correlation functions of a modified theory with the action $S_J[\eta]$:
\begin{equation}
S[\eta] \quad\longrightarrow\quad
S_J[\eta] \equiv S[\eta] + J_x \eta^x \,.
\label{eqn:SJdef}
\end{equation}
The partition function $Z[J]$ generates correlation functions for these generalized theories that include non-trivial dependence on the sources. The correlation functions of the \emph{original theory} $S[\eta]$ can be extracted from their generalized counterparts by taking the zero source condition $J(x)=0$:
\begin{equation}
\langle \eta^{x_1} \cdots \eta^{x_n} \rangle_{J=0} = \frac{\int \Dcal\s\eta\, e^{iS[\eta]}\, \eta(x_1) \cdots \eta(x_n)}{\int \Dcal\s\eta\, e^{iS[\eta]}} = \frac{1}{Z[J=0]} (-i)^n \frac{\delta^n Z}{\delta J_{x_1} \cdots \delta J_{x_n}} \bigg|_{J=0} \,.
\label{eqn:CorrJ0}
\end{equation}
Meanwhile, it is useful to work with the source dependent theories, whose correlation functions are given in \cref{eqn:CorrJ}. Their functional dependence on $J$ is key to the off-shell recursion relation.

\subsubsection*{Connected and 1PI Correlation Functions}

It is more convenient to work with $W[J]$ defined in \cref{eqn:ZWdef}, since this is the generating functional for the contributions from the connected diagrams
\begin{equation}
\langle \eta^{x_1} \cdots \eta^{x_n} \rangle_{J,\,\text{conn}} = (-i)^n \frac{\delta^n (iW)}{\delta J_{x_1} \cdots \delta J_{x_n}} \,.
\label{eqn:connJ}
\end{equation}
The one-particle-irreducible (1PI) effective action $\Gamma[\phi]$ is then defined as a Legendre transform of $W[J]$:
\begin{equation}
\phi^x [J] \equiv \frac{\delta W}{\delta J_x}
\qquad\Longrightarrow\qquad
\Gamma [\phi] \equiv W\big[ J[\phi] \big] - \phi^x J_x[\phi] \,.
\label{eqn:phiGammadef}
\end{equation}
To implement the Legendre transform, one introduces a new set of variables, the set of fields $\phi(x)$ \emph{defined} as in \cref{eqn:phiGammadef}. By construction, these are ``conjugate variables'' to the source fields $J(x)$, in that there is an invertible map between them determined by
\cref{eqn:phiGammadef}:
\begin{equation}
J(x) \quad\longleftrightarrow\quad \phi(x) \,.
\label{eqn:Jphimap}
\end{equation}

We emphasize that the fields $\phi(x)$  are \emph{not} the scalar fields $\eta(x)$ of the theory. However, making use of the $n=1$ case of \cref{eqn:connJ}, one derives a relation between $\phi(x)$ and $\eta(x)$;
$\phi(x)$ are the $J$-dependent quantum vacuum expectation values (vev) of the fields $\eta(x)$:
\begin{equation}
\phi^x [J] \equiv \frac{\delta W}{\delta J_x} = \langle \eta^x \rangle_J \,.
\end{equation}
Some other relations also follow from the general properties of the Legendre transform
\begin{equation}
\frac{\delta\Gamma}{\delta\phi^x} = - J_x \,,\qquad\text{and}\qquad
\frac{\delta^2 (i\Gamma)}{\delta\phi^x \delta\phi^y} = \left[ \frac{\delta^2 (iW)}{\delta J_x \delta J_y} \right]^{-1} \,.
\label{eqn:Gamma12Legendre}
\end{equation}

It is well known that $i\Gamma[\phi]$ is the generating functional of the $J$-dependent 1PI correlation functions
\begin{equation}
\langle \eta(x_1) \cdots \eta(x_n) \rangle_{J,\,\text{1PI}} = \frac{\delta^n (i\Gamma)}{\delta\phi(x_1) \cdots \delta\phi(x_n)} \qquad\text{for}\qquad
n \ge 3 \,.
\label{eqn:1PIJdef}
\end{equation}
The 1PI correlation functions for the original theory are then recovered by taking the zero source condition $J(x)=0$. Through the one-to-one map in \cref{eqn:Jphimap}, this corresponds to evaluating the right-hand side of \cref{eqn:1PIJdef} at a specific choice of $\phi(x)$:
\begin{equation}
J(x)=0 \quad\longleftrightarrow\quad
\phi(x) |_{J=0} = \phi_v(x) \equiv \langle \eta^x \rangle_{J=0} \,.
\label{eqn:phivdef}
\end{equation}
We see that $\phi_v(x)$ is the quantum vev of the fields $\eta(x)$ for the original theory. According to \cref{eqn:Gamma12Legendre}, it satisfies the condition
\begin{equation}
\frac{\delta\Gamma}{\delta\phi^x} \bigg|_{\phi=\phi_v} = 0 \,.
\label{eqn:phiv}
\end{equation}

The 1PI effective action $\Gamma[\phi]$ can be computed as a series of ``1PI diagrams,'' which are diagrams with the property that they cannot be separated into two disconnected parts that each contains a nonzero number of external legs by cutting a single internal leg. One subtle case is that diagrams with tadpoles can be consistent with the 1PI requirement; cutting off the tadpole could separate the diagram into two disconnected parts, but the part including the tadpole does not contain any external legs. Therefore, when computing the 1PI effective action $\Gamma[\phi]$ diagrammatically, one must include diagrams with tadpoles (when they are nonzero), see \eg~\cite{Manohar:2020nzp}.

\subsection{Amplitudes From Correlation Functions}
\label{subsec:Amplitudes}

To compute the amplitudes from the correlation functions, we first define the on-shell momenta. For this purpose, we study the connected two-point functions, namely the propagators:
\begin{equation}
D^{xy}[J] \equiv \langle \eta^x \eta^y \rangle_{J,\,\text{conn}} = - \frac{\delta^2 (iW)}{\delta J_x \delta J_y} = - \left[ \frac{\delta^2 (i\Gamma)}{\delta\phi^x \delta\phi^y} \right]^{-1} \,,
\label{eqn:Propdef}
\end{equation}
where the second-to-last expression comes from \cref{eqn:connJ}, while the last equality is due to the property of the Legendre transform in \cref{eqn:Gamma12Legendre}. Again, this is the propagator for the $J$-dependent theory $S_J[\eta]=S[\eta]+J_x\eta^x$. Taking the zero source condition, $J(x)=0$ or equivalently $\phi(x)=\phi_v(x)$, recovers the propagator of the original theory $S[\eta]$. Its momentum space form is the familiar one:
\begin{equation}
\int \dd^4 x_1 \dd^4 x_2\, e^{i p_1 x_1} e^{i p_2 x_2}\, D^{x_1x_2}[J=0]
= (2\pi)^4 \delta^4(p_1+p_2)\, \Delta(p_1) \,,
\label{eqn:DJ0Fpdelta}
\end{equation}
with
\begin{equation}
\Delta(p) = \frac{i R_\eta}{p^2 - m_p^2 + i\epsilon} + \text{regular} \,,
\label{eqn:Fpdef}
\end{equation}
where $m_p$ denotes the pole mass of the particle and $R_\eta$ denotes the residue. Using \cref{eqn:Propdef}, one can write \cref{eqn:DJ0Fpdelta} alternatively as
\begin{equation}
\int \dd^4 x_1\, e^{ip(x_1-x_2)}\,
\frac{\delta^2 \Gamma}{\delta\phi^{x_1} \delta\phi^{x_2}} \bigg|_{\phi=\phi_v}
= \int \dd^4 x_1\, e^{ip(x_1-x_2)}\, iD_{x_1x_2}^{-1}[J=0]
= \frac{i}{\Delta(p)} \,.
\label{eqn:DInvJ0FpGamma}
\end{equation}

Note that \cref{eqn:Propdef} is the fully connected two-point function, or the full interacting propagator. Specifically, if we denote the 1PI two-point function as $-i\Sigma(p^2)$, we have
\begin{align}
\Delta(p) &= \frac{i}{p^2 - m^2 + i\epsilon} + \frac{i}{p^2 - m^2 + i\epsilon}\, \big[ -i\Sigma(p^2) \big]\, \frac{i}{p^2 - m^2 + i\epsilon} + \cdots
\notag\\[5pt]
&= \frac{i}{p^2 - m^2 - \Sigma(p^2) + i\epsilon} \,,
\end{align}
where $m^2$ is the tree-level mass parameter, and the pole mass $m_p^2$ is determined by the condition $m_p^2 = m^2 + \Re \Sigma(m_p^2)$.

\subsubsection*{On-shell Condition}

A momentum $p^\mu$ is said to be on-shell when it sits on the pole of the propagator
\begin{equation}
\frac{1}{\Delta(\pon)} = 0
\qquad\Longrightarrow\qquad
\pon^2 = m_p^2 \,,
\label{eqn:OnShellp}
\end{equation}
where we are introducing the notation $\pon$ to denote on-shell momenta. Using \cref{eqn:DInvJ0FpGamma}, we can equivalently state the on-shell condition as
\begin{equation}
\int \dd^4 x_1\, e^{i \pon x_1}\,
\frac{\delta^2\Gamma}{\delta\phi^{x_1} \delta\phi^{x_2}} \bigg|_{\phi=\phi_v} = 0 \,.
\label{eqn:OnShellpGamma}\vspace{3mm}
\end{equation}

\clearpage

\subsubsection*{Amplitudes From LSZ and External Wavefunctions}

To compute the amplitudes following the LSZ prescription \cite{Lehmann:1954rq, Lehmann:1957zz}, one can first compute the $J$-dependent amputated correlation functions
\begin{equation}
-i\Mcal_{x_1 \cdots x_n}[J] \equiv \left( D_{x_1 y_1}^{-1} \right) \cdots \left( D_{x_n y_n}^{-1} \right) \langle \eta^{y_1} \cdots \eta^{y_n} \rangle_{J,\,\text{conn}} \,.
\label{eqn:Mdef}
\end{equation}
Then the momentum space amplitudes $\Amp$ follow by evaluating $\Mcal$ at $J=0$, taking a Fourier transform, and including the appropriate residue factors for the external legs:
\begin{align}
(2\pi)^4 \delta^4(p_1 +\cdots + p_n)\, &i\Amp \left( p_1, \cdots, p_n \right) \notag\\[8pt]
&= (R_\eta^{1/2})^n \int \left[ \prod_{i=1}^n \dd^4x_i\, e^{i p_i x_i} \right] \Bigl( -i\Mcal_{x_1 \cdots x_n}|_{J=0} \Bigr) \,.
\label{eqn:AmpdefFourier}
\end{align}
This defines general amplitudes for off-shell momenta $p_i^2 \neq m_{p,i}^2$.
The on-shell amplitudes (the usual $S$-matrix elements) are then given by taking all external momenta to be on shell.

It is convenient to introduce the external wavefunction\footnote{In general, the external wavefunctions are $\psi_i = \mel**{0}{\eta^{x_i}}{p_i, h_i, \cdots}_{J=0}$, which represent the overlap of the fields with the $i^\text{th}$ external states of given momentum $p_i$, helicity $h_i$, \etc\ For an external scalar, it has the form in \cref{eqn:psixScalar}, whereas for a gauge boson, it would also include a polarization vector, \ie, $\psi_i = \epsilon_{h_i}^{\mu_i}(p_i)\, e^{i p_i x_i}$.}
\begin{equation}
\psi^x(p) = R_\eta^{1/2}\, e^{i p x} \,,
\label{eqn:psixScalar}
\end{equation}
which is an eigenstate of the inverse propagator (\cf\ \cref{eqn:DInvJ0FpGamma}):
\begin{equation}
\frac{\delta^2\Gamma}{\delta\phi^{x_1} \delta\phi^{x_2}} \bigg|_{\phi=\phi_v}\, \psi^{x_2}(p)
= \frac{i}{\Delta(p)}\, \psi^{x_1}(p) \,.
\label{eqn:psixEigen}
\end{equation}
Note that when the momentum is on-shell, the eigenvalue vanishes
\begin{equation}
\frac{\delta^2\Gamma}{\delta\phi^{x_1} \delta\phi^{x_2}} \bigg|_{\phi=\phi_v}\, \psi^{x_2}(\pon) = 0 \,.
\label{eqn:psixOnShellp}
\end{equation}
With the external wavefunctions, we can write the LSZ formula in \cref{eqn:AmpdefFourier} more concisely as
\begin{equation}
(2\pi)^4 \delta^4(p_1 +\cdots + p_n)\, i\Amp \left( p_1, \cdots, p_n \right)
= \big[ \psi^{x_1}(p_1) \cdots \psi^{x_n}(p_n) \big]\, \bigl( -i\Mcal_{x_1 \cdots x_n} |_{J=0} \bigr) \,,
\label{eqn:Ampdef}
\end{equation}
compare analogous equations in \cite{Cheung:2021yog,Alonso:2022ffe}. We emphasize here that $\Amp \left( p_1, \cdots, p_n \right)$ defines a generalized momentum space amplitude where the external momenta can be off-shell.

\subsubsection*{Computing Amputated Correlation Functions}

In order to compute amplitudes, \cref{eqn:Ampdef} implies that we can focus on calculating the amputated correlation functions $-i\Mcal_{x_1 \cdots x_n}[J]$ defined in \cref{eqn:Mdef}. These can be obtained by gluing together the 1PI correlation functions (\cref{eqn:1PIJdef}) using the propagators (\cref{eqn:Propdef}). As discussed above, these two types of building blocks for $-i\Mcal_{x_1 \cdots x_n}$ are both conveniently expressed in terms of the 1PI effective action. Concretely, the three-point amputated correlation function can be expressed as
\begin{equation}
-i\Mcal_{x_1 x_2 x_3} = \frac{\delta^3 (i\Gamma)}{\delta\phi^{x_1} \delta\phi^{x_2} \delta\phi^{x_3}} \,,
\label{eqn:M3}
\end{equation}
while at four-points we have
\begin{align}
-i\Mcal_{x_1 x_2 x_3 x_4} &= \frac{\delta^4 (i\Gamma)}{\delta\phi^{x_1} \delta\phi^{x_2} \delta\phi^{x_3} \delta\phi^{x_4}} + \frac{\delta^3 (i\Gamma)}{\delta\phi^{x_1} \delta\phi^{x_2} \delta\phi^y} D^{yz} \frac{\delta^3 (i\Gamma)}{\delta\phi^z \delta\phi^{x_3} \delta\phi^{x_4}}
\notag\\[5pt]
&\hspace{-10pt}
+ \frac{\delta^3 (i\Gamma)}{\delta\phi^{x_1} \delta\phi^{x_3} \delta\phi^y} D^{yz} \frac{\delta^3 (i\Gamma)}{\delta\phi^z \delta\phi^{x_2} \delta\phi^{x_4}}
+ \frac{\delta^3 (i\Gamma)}{\delta\phi^{x_1} \delta\phi^{x_4} \delta\phi^y} D^{yz} \frac{\delta^3 (i\Gamma)}{\delta\phi^z \delta\phi^{x_2} \delta\phi^{x_3}} \,.
\label{eqn:M4}
\end{align}
Similar expressions can be worked out for higher-point functions. As we will explain next, they can more efficiently be built recursively out of lower-point functions.

\subsection{Recursion Relation for Amplitudes}
\label{subsec:Recursion}

We now explain how to derive higher point generalizations of \cref{eqn:M3,eqn:M4} recursively. For convenience, we introduce notation for the following combination of the three-point function and the propagator:
\begin{equation}
\Chris_{x_1x_2}^y \equiv i\Mcal_{x_1 x_2 z}\, D^{zy} \,.
\label{eqn:Chrisdef}
\end{equation}
With this, one can rewrite \cref{eqn:M4} as
\begin{equation}
\Mcal_{x_1 x_2 x_3 x_4} = \frac{\delta}{\delta\phi^{x_4}} \Mcal_{x_1 x_2 x_3}
- \Chris_{x_4 x_1}^y \Mcal_{y x_2 x_3}
- \Chris_{x_4 x_2}^y \Mcal_{x_1 y x_3}
- \Chris_{x_4 x_3}^y \Mcal_{x_1 x_2 y} \,.
\label{eqn:M4Recursion}
\end{equation}
This way of writing the four-point function exposes a relation to the three-point function. We will now argue that this pattern persists to any number of external legs as a recursion relation of the form \cite{Cohen:2022uuw}
\begin{equation}
\Mcal_{x_1 \cdots x_n x_{n+1}} = \frac{\delta}{\delta\phi^{x_{n+1}}}\, \Mcal_{x_1 \cdots x_n}
- \sum_{i=1}^n \Chris_{x_{n+1} x_i}^y\, \Mcal_{x_1 \cdots \hat{x}_i y \cdots x_n} \,,
\label{eqn:Recursion}
\end{equation}
where a hat denotes the absence of an index in the sequence. Note that the form of \cref{eqn:Recursion} is suggestive of a covariant derivative where $G$ is the connection; we return to this point in \cref{sec:Geometry}.

To derive this recursion relation, we first use the definition of $\Mcal_{x_1 \cdots x_n}$ in \cref{eqn:Mdef} together with \cref{eqn:connJ} to obtain
\begin{equation}
-i\Mcal_{x_1 \cdots x_n} = D_{x_1 y_1}^{-1} \cdots D_{x_n y_n}^{-1}
(-i)^n \frac{\delta^n (iW)}{\delta J_{y_1} \cdots \delta J_{y_n}} \,.
\label{eqn:McalDW}
\end{equation}
This implies the following relation between $\Mcal_{x_1 \cdots x_n x_{n+1}}$ and $\Mcal_{x_1 \cdots x_n}$:
\begin{align}
\Mcal_{x_1 \cdots x_n x_{n+1}} &= D_{x_1 y_1}^{-1} \cdots D_{x_n y_n}^{-1} D_{x_{n+1} y_{n+1}}^{-1}
\left(-i\right)\frac{\delta}{\delta J_{y_{n+1}}}\,
D^{y_1 z_1} \cdots D^{y_n z_n} \Mcal_{z_1 \cdots z_n}
\notag\\[5pt]
&= D_{x_1 y_1}^{-1} \cdots D_{x_n y_n}^{-1}\,
\frac{\delta}{\delta\phi^{x_{n+1}}}\,
D^{y_1 z_1} \cdots D^{y_n z_n} \Mcal_{z_1 \cdots z_n} \,,
\label{eqn:Mnplus1D}
\end{align}
where we have used \cref{eqn:phiGammadef,eqn:Propdef} to obtain the second line. We can simplify this expression using the commutator between the functional derivative $\frac{\delta}{\delta\phi^{x_{n+1}}}$ and the propagators. Using \cref{eqn:Propdef} again, together with \cref{eqn:M3} and the definition in \cref{eqn:Chrisdef}, we get
\begin{align}
\comm{\frac{\delta}{\delta\phi^{x_{n+1}}}}{D^{y_i z_i}}
&= \left(\frac{\delta}{\delta\phi^{x_{n+1}}}\, D^{y_i z_i} \right)
= - \left( \frac{\delta}{\delta\phi^{x_{n+1}}} \left[ \frac{\delta^2 (i\Gamma)}{\delta\phi^{y_i} \delta\phi^{z_i}} \right]^{-1} \right)
\notag\\[8pt]
&= \left[ \frac{\delta^2 (i\Gamma)}{\delta\phi^{y_i} \delta\phi^u} \right]^{-1}
\frac{\delta^3 (i\Gamma)}{\delta\phi^u \delta\phi^{x_{n+1}} \delta\phi^v}
\left[ \frac{\delta^2 (i\Gamma)}{\delta\phi^v \delta\phi^{z_i}} \right]^{-1}
\notag\\[10pt]
&= - D^{y_i u}\, i\Mcal_{x_{n+1} u v}\, D^{v z_i} = - D^{y_iu}\, \Chris_{x_{n+1} u}^{z_i} \,.
\label{eqn:commutator}
\end{align}
Using this repeatedly, we obtain the following relation
\begin{align}
D_{x_1 y_1}^{-1} \cdots D_{x_n y_n}^{-1}\,
&\frac{\delta}{\delta\phi^{x_{n+1}}}\,
D^{y_1 z_1} \cdots D^{y_n z_n}
\notag\\[6pt]
&= \delta_{x_1}^{z_1} \cdots \delta_{x_n}^{z_n}\, \frac{\delta}{\delta\phi^{x_{n+1}}}
- \sum_{i=1}^n \left( \delta_{x_1}^{z_1} \cdots \hat\delta_{x_i}^{z_i} \cdots \delta_{x_n}^{z_n} \right) \Chris_{x_{n+1} x_i}^{z_i} \,,
\end{align}
where $\delta_x^z = \frac{\delta\phi^z}{\delta\phi^x} = \delta^4(z-x)$ and the hat indicates the absence of a quantity in the sequence as before. With this relation, \cref{eqn:Mnplus1D} simplifies to the recursion relation in \cref{eqn:Recursion}. Note that no step in this derivation relied on any reference to perturbation theory. Therefore, the recursion relation for $\Mcal_{x_1 \cdots x_n}$ in \cref{eqn:Recursion} holds to all loop orders.

\subsubsection{Diagrammatic Derivation}
\label{subsubsec:Diagrammatic}

The above derivation of the recursion relation \cref{eqn:Recursion} is purely algebraic. To provide a more intuitive perspective, we present a diagrammatic derivation in this section, which repeats the argument given in \cite{Cohen:2022uuw} with more details.

Consider the diagrammatic representation of the amputated correlation functions $-i\Mcal_{x_1 \cdots x_n}$. They can be obtained by gluing together the 1PI vertices with the full propagators; both ingredients are conveniently expressed in terms of the 1PI effective action, as shown in \cref{eqn:1PIJdef,eqn:Propdef}. Here we recap the dictionary between diagram components and algebraic factors for our convenience:
\begin{subequations}\label{eqn:Dictionary}
\begin{align}
k\text{-point 1PI vertices}: &\qquad   \frac{\delta^k (i\Gamma)}{\delta\phi^{x_1} \cdots \delta\phi^{x_k}} \, ,\quad k \geq 3 \,, \label{eqn:VkGamma} \\[5pt]
\text{full propagators}: &\qquad   D^{xy} = - \left[ \frac{\delta^2 (i\Gamma)}{\delta\phi^x \delta\phi^y} \right]^{-1} \,. \label{eqn:DGamma}
\end{align}
\end{subequations}%
As usual, we group all the contributing Feynman diagrams into different ``gluing topologies,'' which characterize all possible ways of gluing together 1PI vertices. For example, at $n=3$ there is a unique gluing topology:
\begin{equation}
-i\Mcal_{x_1 x_2 x_3} = \mathord{
  \begin{tikzpicture}[baseline=-0.65ex]
  \node[vertexstyle] (centralnode) at (0,0) {\footnotesize 1PI};
  \draw (centralnode) -- (90:0.8) node[shift={(90:0.25)}] {\scriptsize $x_1$};
  \draw (centralnode) -- (210:0.8) node[shift={(210:0.25)}] {\scriptsize $x_2$};
  \draw (centralnode) -- (-30:0.8) node[shift={(-30:0.25)}] {\scriptsize $x_3$};
  \end{tikzpicture}} \,.
\end{equation}
This corresponds to the single term in \cref{eqn:M3}. At $n=4$, there are four distinct gluing topologies, each corresponding to a term in \cref{eqn:M4}:
\begin{align}
-i\Mcal_{x_1 x_2 x_3 x_4} &= \mathord{
  \begin{tikzpicture}[baseline=-0.65ex]
  \node[vertexstyle] (centralnode) at (0,0) {\footnotesize 1PI};
  \draw (centralnode) -- (135:0.8) node[shift={(135:0.25)}] {\scriptsize $x_1$};
  \draw (centralnode) -- (225:0.8) node[shift={(225:0.25)}] {\scriptsize $x_2$};
  \draw (centralnode) -- (-45:0.8) node[shift={(-45:0.25)}] {\scriptsize $x_3$};
  \draw (centralnode) -- (45:0.8) node[shift={(45:0.25)}] {\scriptsize $x_4$};
  \end{tikzpicture}}
+ \mathord{
  \begin{tikzpicture}[baseline=-0.65ex]
  \node[vertexstyle] (node1) at (-0.7,0) {\footnotesize 1PI};
  \node[vertexstyle] (node2) at (0.7,0) {\footnotesize 1PI};
  \draw (node1) -- +(135:0.8) node[shift={(135:0.25)}] {\scriptsize $x_1$};
  \draw (node1) -- +(225:0.8) node[shift={(225:0.25)}] {\scriptsize $x_2$};
  \draw (node2) -- +(-45:0.8) node[shift={(-45:0.25)}] {\scriptsize $x_3$};
  \draw (node2) -- +(45:0.8) node[shift={(45:0.25)}] {\scriptsize $x_4$};
  \draw[propstyle] (node1) -- node[pos=0.2,shift={(270:0.2)}] {\scriptsize $y$} node[pos=0.8,shift={(270:0.2)}] {\scriptsize $z$} (node2);
  \end{tikzpicture}}
+ \mathord{
  \begin{tikzpicture}[baseline=-0.65ex]
  \node[vertexstyle] (node1) at (-0.7,0) {\footnotesize 1PI};
  \node[vertexstyle] (node2) at (0.7,0) {\footnotesize 1PI};
  \draw (node1) -- +(135:0.8) node[shift={(135:0.25)}] {\scriptsize $x_1$};
  \draw (node1) -- +(225:0.8) node[shift={(225:0.25)}] {\scriptsize $x_3$};
  \draw (node2) -- +(-45:0.8) node[shift={(-45:0.25)}] {\scriptsize $x_2$};
  \draw (node2) -- +(45:0.8) node[shift={(45:0.25)}] {\scriptsize $x_4$};
  \draw[propstyle] (node1) -- node[pos=0.2,shift={(270:0.2)}] {\scriptsize $y$} node[pos=0.8,shift={(270:0.2)}] {\scriptsize $z$} (node2);
  \end{tikzpicture}}
\notag\\[5pt]
&\qquad
+ \mathord{
  \begin{tikzpicture}[baseline=-0.65ex]
  \node[vertexstyle] (node1) at (-0.7,0) {\footnotesize 1PI};
  \node[vertexstyle] (node2) at (0.7,0) {\footnotesize 1PI};
  \draw (node1) -- +(135:0.8) node[shift={(135:0.25)}] {\scriptsize $x_1$};
  \draw (node1) -- +(225:0.8) node[shift={(225:0.25)}] {\scriptsize $x_4$};
  \draw (node2) -- +(-45:0.8) node[shift={(-45:0.25)}] {\scriptsize $x_3$};
  \draw (node2) -- +(45:0.8) node[shift={(45:0.25)}] {\scriptsize $x_2$};
  \draw[propstyle] (node1) -- node[pos=0.2,shift={(270:0.2)}] {\scriptsize $y$} node[pos=0.8,shift={(270:0.2)}] {\scriptsize $z$} (node2);
  \end{tikzpicture}}
\,.
\label{eqn:M4Diagram}
\end{align}

Now let us consider the gluing topologies for $-i\Mcal_{x_1 \cdots x_n}$ and $-i\Mcal_{x_1 \cdots x_n x_{n+1}}$, with $n\ge3$. The latter has one more leg, $x_{n+1}$, and hence receives contributions from more gluing topologies. We can examine each of them, paying attention to where the extra leg $x_{n+1}$ is attached. In this way, for each gluing topology $T_{n+1}$ of $-i\Mcal_{x_1 \cdots x_n x_{n+1}}$, one can first identify a corresponding gluing topology $T_n$ of $-i\Mcal_{x_1 \cdots x_n}$, and then figure out how one can calculate $T_{n+1}$ from $T_n$. 

Let us elaborate this procedure in detail. Specifically, there are three scenarios for the position of the leg $x_{n+1}$ in $T_{n+1}$:
\begin{enumerate}
\item $x_{n+1}$ is part of a four- (or higher-) point 1PI vertex in the gluing topology $T_{n+1}$. In this case, if one removes $x_{n+1}$, the 1PI vertex that it is attaching to will remain as a 1PI vertex, and $T_{n+1}$ will become a gluing topology $T_n$ for $-i\Mcal_{x_1 \cdots x_n}$. Diagrammatically, one identifies the corresponding $T_n$ from $T_{n+1}$ as
\begin{align}
T_{n+1} = \;\;\mathord{
  \begin{tikzpicture}[baseline=-0.65ex]
  \node[vertexstyle] (centralnode) at (0,0) {\footnotesize 1PI};
  \node[vertexstyle,gray] (node1) at (180:1.2) {\footnotesize 1PI};
  \node[vertexstyle,gray] (node2) at (135:1.2) {\footnotesize 1PI};
  \draw[propstyle] (centralnode) -- node[pos=0.2,shift={(270:0.2)}] {\scriptsize $y_1$} (node1);
  \draw[propstyle] (centralnode) -- node[pos=0.2,shift={(45:0.2)}] {\scriptsize $y_m$} (node2);
  \draw[dotstyle,gray] ([shift=(90:\dotrad)]node1) arc (90:270:\dotrad);
  \draw[dotstyle,gray] ([shift=(45:\dotrad)]node2) arc (45:225:\dotrad);
  \draw[dotstyle] ([shift=(140:\dotrad)]centralnode) arc (140:175:\dotrad);
  \draw (centralnode) -- (70:0.7) node[shift={(70:0.2)}] {\scriptsize $y_{m+1}$};
  \draw (centralnode) -- (25:0.7) node[shift={(25:0.2)}] {\scriptsize $y_{k}$};
  \draw[dotstyle] ([shift=(30:\dotrad)]centralnode) arc (30:65:\dotrad);
  \draw (centralnode) -- (0:0.7) node[shift={(0:0.4)}] {\scriptsize $x_{n+1}$};
  \end{tikzpicture}}
\qquad\Longrightarrow\qquad
T_n = \;\;\mathord{
  \begin{tikzpicture}[baseline=-0.65ex]
  \node[vertexstyle] (centralnode) at (0,0) {\footnotesize 1PI};
  \node[vertexstyle,gray] (node1) at (180:1.2) {\footnotesize 1PI};
  \node[vertexstyle,gray] (node2) at (135:1.2) {\footnotesize 1PI};
  \draw[propstyle] (centralnode) -- node[pos=0.2,shift={(270:0.2)}] {\scriptsize $y_1$} (node1);
  \draw[propstyle] (centralnode) -- node[pos=0.2,shift={(45:0.2)}] {\scriptsize $y_m$} (node2);
  \draw[dotstyle,gray] ([shift=(90:\dotrad)]node1) arc (90:270:\dotrad);
  \draw[dotstyle,gray] ([shift=(45:\dotrad)]node2) arc (45:225:\dotrad);
  \draw[dotstyle] ([shift=(140:\dotrad)]centralnode) arc (140:175:\dotrad);
  \draw (centralnode) -- (70:0.7) node[shift={(70:0.2)}] {\scriptsize $y_{m+1}$};
  \draw (centralnode) -- (25:0.7) node[shift={(25:0.2)}] {\scriptsize $y_{k}$};
  \draw[dotstyle] ([shift=(30:\dotrad)]centralnode) arc (30:65:\dotrad);
  \end{tikzpicture}}\;\; \,.
\end{align}
Next, from the dictionary in \cref{eqn:VkGamma}, we see that $T_{n+1}$ can be calculated from $T_n$ by taking a functional derivative $\frac{\delta}{\delta\phi^{x_{n+1}}}$ of the corresponding vertex factor in $T_n$, because
\begin{equation}
\frac{\delta^{k+1}(i\Gamma)}{\delta\phi^{y_1} \cdots \delta\phi^{y_k} \delta\phi^{x_{n+1}}} =
\frac{\delta}{\delta\phi^{x_{n+1}}} \left( \frac{\delta^k(i\Gamma)}{\delta\phi^{y_1} \cdots \delta\phi^{y_k}} \right) \,.
\end{equation}
\item $x_{n+1}$ is part of a three-point 1PI vertex in the gluing topology $T_{n+1}$, and none of the other two lines from this 1PI vertex is a leg. In this case, one can remove $x_{n+1}$ by replacing the three-point 1PI vertex with a propagator, and thus obtain a gluing topology $T_n$ for $-i\Mcal_{x_1 \cdots x_n}$. Diagrammatically, one identifies the corresponding $T_n$ from $T_{n+1}$ as
\begin{align}
T_{n+1} = \;\;\mathord{
  \begin{tikzpicture}[baseline=-0.65ex]
  \node[vertexstyle] (node1) at (-1.4,-0.2) {\footnotesize 1PI};
  \node[vertexstyle] (node2) at (1.4,-0.2) {\footnotesize 1PI};
  \node[vertexstyle] (centralnode) at (0,0.2) {\footnotesize 1PI};
  \draw[propstyle] (node1) -- node[pos=0.2,shift={(285:0.2)}] {\scriptsize $y_1$} node[pos=0.8,shift={(285:0.2)}] {\scriptsize $z_1$} (centralnode);
  \draw[propstyle] (node2) -- node[pos=0.2,shift={(255:0.2)}] {\scriptsize $y_2$} node[pos=0.8,shift={(255:0.2)}] {\scriptsize $z_2$} (centralnode);
  \draw[dotstyle] ([shift=(90:\dotrad)]node1) arc (90:270:\dotrad);
  \draw[dotstyle] ([shift=(-90:\dotrad)]node2) arc (-90:90:\dotrad);
  \draw (centralnode) -- (90:0.9) node[shift={(90:0.2)}] {\scriptsize $x_{n+1}$};
  \end{tikzpicture}}
\qquad\Longrightarrow\qquad
T_n = \;\;\mathord{
  \begin{tikzpicture}[baseline=-0.65ex]
  \node[vertexstyle] (node1) at (-0.8,0) {\footnotesize 1PI};
  \node[vertexstyle] (node2) at (0.8,0) {\footnotesize 1PI};
  \draw[propstyle] (node1) -- node[pos=0.2,shift={(270:0.2)}] {\scriptsize $y_1$} node[pos=0.8,shift={(270:0.2)}] {\scriptsize $y_2$} (node2);
  \draw[dotstyle] ([shift=(90:\dotrad)]node1) arc (90:270:\dotrad);
  \draw[dotstyle] ([shift=(-90:\dotrad)]node2) arc (-90:90:\dotrad);
  \end{tikzpicture}}\;\; \,.
\end{align}
Next, from the dictionary in \cref{eqn:DGamma}, we see that $T_{n+1}$ can be calculated from $T_n$ by taking a functional derivative $\frac{\delta}{\delta\phi^{x_{n+1}}}$ of the corresponding propagator factor in $T_n$, because
\begin{equation}
D^{y_1 z_1} \frac{\delta^3 (i\Gamma)}{\delta\phi^{z_1} \delta\phi^{x_{n+1}} \delta\phi^{z_2}} D^{z_2 y_2}
= \frac{\delta}{\delta\phi^{x_{n+1}}} \left( D^{y_1 y_2} \right) \,. 
\end{equation}
\item $x_{n+1}$ is part of a three-point 1PI vertex in the gluing topology $T_{n+1}$, and one of the other two lines from this 1PI vertex is a leg $x_i$. (For $n\ge3$, one cannot have both the other two lines being legs.) In this case, one can remove $x_{n+1}$ by cutting off the three-point 1PI vertex from the diagram and relabeling the leg from the cut as $x_i$ to get a gluing topology $T_n$ for $-i\Mcal_{x_1 \cdots x_n}$. Diagrammatically, one identifies the corresponding $T_n$ from $T_{n+1}$ as
\begin{align}
T_{n+1} = \;\;\mathord{
  \begin{tikzpicture}[baseline=-0.65ex]
  \node[vertexstyle,fill=gray] (centralnode) at (0,0) {\footnotesize $\Mcal$};
  \node[vertexstyle] (newnode) at (-1.2,0) {\footnotesize 1PI};
  \draw[propstyle] (centralnode) -- node[pos=0.2,shift={(270:0.2)}] {\scriptsize $y$} node[pos=0.8,shift={(270:0.17)}] {\scriptsize $z$} (newnode);
  \draw (centralnode) -- (45:0.7) node[shift={(45:0.2)}] {\scriptsize $x_{1}$};
  \draw (centralnode) -- (-45:0.7) node[shift={(-45:0.2)}] {\scriptsize $x_{n}$};
  \draw (newnode) -- +(135:0.7) node[shift={(135:0.2)}] {\scriptsize $x_{n+1}$};
  \draw (newnode) -- +(-135:0.7) node[shift={(-135:0.2)}] {\scriptsize $x_{i}$};
  \draw[dotstyle] ([shift=(50:\dotradM)]centralnode) arc (50:175:\dotradM);
  \draw[dotstyle] ([shift=(235:\dotradM)]centralnode) arc (235:310:\dotradM);
  \end{tikzpicture}}
\qquad\Longrightarrow\qquad
T_n = \;\;\mathord{
  \begin{tikzpicture}[baseline=-0.65ex]
  \node[vertexstyle,fill=gray] (centralnode) at (0,0) {\footnotesize $\Mcal$};
  \draw (centralnode) -- (180:0.7) node[shift={(180:0.2)}] {\scriptsize $x_{i}$};
  \draw (centralnode) -- (45:0.7) node[shift={(45:0.2)}] {\scriptsize $x_{1}$};
  \draw (centralnode) -- (-45:0.7) node[shift={(-45:0.2)}] {\scriptsize $x_{n}$};
  \draw[dotstyle] ([shift=(50:\dotradM)]centralnode) arc (50:175:\dotradM);
  \draw[dotstyle] ([shift=(185:\dotradM)]centralnode) arc (185:310:\dotradM);
  \end{tikzpicture}} \,.
\end{align}
Next, from the definition in \cref{eqn:Chrisdef}, we see that $T_{n+1}$ can be calculated from $T_n$ by first taking the replacement $x_i \to y$ and then contracting with the factor $-\Chris_{x_{n+1} x_i}^y$, because
\begin{equation}
\frac{\delta^3 (i\Gamma)}{\delta\phi^{x_{n+1}} \delta\phi^{x_i} \delta\phi^z}\, D^{zy}\, \Mcal_{x_1 \cdots \hat{x}_i y \cdots x_n} = - \Chris_{x_{n+1} x_i}^y\, \Mcal_{x_1 \cdots \hat{x}_i y \cdots x_n} \,.
\end{equation}
\end{enumerate}
In summary, scenarios 1 and 2 together gives the functional derivative term in \cref{eqn:Recursion}, and scenario 3 gives us the terms involving the contraction with $\Chris_{x_{n+1}x_i}^y$. This completes the diagrammatic proof of the recursion relation in \cref{eqn:Recursion}.

\subsubsection{Connection with Berends-Giele Recursion Relation}
\label{subsubsec:BerendsGiele}

Since it involves off-shell building blocks, the recursion relation in \cref{eqn:Recursion} can be related to the Berends-Giele off-shell recursion relation for computing the amplitudes \cite{Berends:1987me, Brown:1992ay, Monteiro:2011pc} as we now explain. First, all the amputated correlation functions (and therefore the amplitudes) are encoded in the functional relation $\phi^x[J]$. Specifically, using our definition of the field $\phi^x$ in \cref{eqn:phiGammadef}, we can rewrite \cref{eqn:McalDW} as
\begin{equation}
-i\Mcal_{x_1 \cdots x_n} = \left( D_{x_1 y_1}^{-1} \right) \cdots \left( D_{x_n y_n}^{-1} \right) (-i)^{n-1} \frac{\delta^{n-1}\phi^{y_1}}{\delta J_{y_2} \cdots \delta J_{y_n}} \,.
\end{equation}
Note also from \cref{eqn:Propdef} that
\begin{equation}
iD_{x_iy_i} = \frac{\delta\phi^{x_i}}{\delta J_{y_i}} \,.
\end{equation} 
Therefore, by rearranging terms and evaluating them at $J=0$, we can obtain the relation between $\Mcal_{x_1 \cdots x_n} |_{J=0}$ and the Taylor expansion coefficients of $\phi^x[J]$ at $J=0$:
\begin{equation}
\frac{\delta^{n-1}\phi^{y_1}}{\delta J_{y_2} \cdots \delta J_{y_n}} \bigg|_{J=0}
= \bigl(-\Mcal_{x_1 \cdots x_n} |_{J=0} \bigr)
\left( \frac{\delta\phi^{x_1}}{\delta J_{y_1}} \bigg|_{J=0} \right) \cdots
\left( \frac{\delta\phi^{x_n}}{\delta J_{y_n}} \bigg|_{J=0} \right) \,.
\label{eqn:MJ0phiTaylor}
\end{equation}
The Berends-Giele approach \cite{Berends:1987me, Brown:1992ay, Monteiro:2011pc} is to iteratively solve the equation of motion condition
\begin{equation}
\frac{\delta\Gamma}{\delta\phi^x} = - J_x \,,
\end{equation}
to obtain the functional relation $\phi^x[J]$ order by order in $J$. This is computing its Taylor expansion coefficients at $J=0$ in \cref{eqn:MJ0phiTaylor}. One can then obtain $\Mcal_{x_1 \cdots x_n} |_{J=0}$ through \cref{eqn:MJ0phiTaylor}. This is in contrast with our recursion relation in \cref{eqn:Recursion}, which directly constructs $\Mcal_{x_1 \cdots x_n}[J]$ recursively for $J\ne 0$.

%%%%%%%%%%%%%%%%%%%%%%%%%%%%%%%%%%%%%%%%%%%%%%%%%%%%%%%%%%%%%%%%%%%%%%%%%%%%%%%%
\section{Invariance of Amplitudes Under General Field Redefinitions}
\label{sec:Invariance}
%%%%%%%%%%%%%%%%%%%%%%%%%%%%%%%%%%%%%%%%%%%%%%%%%%%%%%%%%%%%%%%%%%%%%%%%%%%%%%%%

In \cref{sec:Recursion}, we reviewed how the $n$-point amplitudes $\Amp \left( p_1, \cdots, p_n \right)$, for both on-shell and off-shell kinematics, can be obtained from the $n$-point amputated correlation functions $\Mcal_{x_1 \cdots x_n}[J]$ using LSZ reduction (\cref{eqn:Ampdef}), and we derived an off-shell recursion relation for $\Mcal_{x_1 \cdots x_n}[J]$ (\cref{eqn:Recursion}). Both of these results are well known; the novelty here is how we organize the terms. As we will show in this section, this organization of the results facilitates a new proof of the invariance of on-shell amplitudes under general field redefinitions, including those involving derivatives. Our results here are complementary to the traditional approach that makes the argument directly from the path integral (see \eg\ Section 6.2 of \cite{Manohar:2018aog}, which we also reproduce in \cref{appsec:Invariance} for completeness).

An important lesson learned from \cref{sec:Recursion} is that the amplitudes $\Amp\left( p_1, \cdots, p_n \right)$ are 

\clearpage

\noindent encoded in a given 1PI effective action $\Gamma[\phi]$:
\begin{equation}
\Gamma[\phi]
\quad\longrightarrow\quad
\Mcal_{x_1 \cdots x_n}[J]
\quad\longrightarrow\quad
\Amp\left( p_1, \cdots, p_n \right) \,.
\label{eqn:GammaToAmp}
\end{equation}
This is independent of the loop order; for a given theory $S[\eta]$, the truncation in terms of loop order only impacts the computation of the 1PI effective action $\Gamma[\phi]$ itself. One can therefore explore the properties of amplitudes by analyzing the behavior of the 1PI effective action.\footnote{We mention that this is the exact same spirit of functional methods for EFT matching calculations (\eg\ \cite{Cohen:2020fcu, Cohen:2022tir}), where the matching of amplitudes are efficiently achieved/guaranteed through the matching of the 1PI effective actions.} In particular, we will make use of the recursion relation in \cref{eqn:Recursion} to prove the following \emph{transformation lemma} in \cref{subsec:Proof}:
\vspace{3mm}
\begin{tcolorbox}
\begin{center}
\begin{minipage}{5.5in}
Define the 1PI effective action $\tGamma[\tphi]$ as a transformation of $\Gamma[\phi]$ that results from substituting in a given analytic functional relation $\phi[\tphi]$:
\begin{equation}
\tGamma [\tphi] = \Gamma \big[ \phi[\tphi] \big] \,.
\label{eqn:GammaRelation}
\end{equation}
Then the amputated correlation functions encoded in these two 1PI effective actions, $\tMcal$ and $\Mcal$ respectively, are related by
\begin{equation}
\tMcal_{x_1 \cdots x_n} = \frac{\delta\phi^{y_1}}{\delta\tphi^{x_1}} \cdots \frac{\delta\phi^{y_n}}{\delta\tphi^{x_n}}\, \Mcal_{y_1 \cdots y_n} + U_{x_1 \cdots x_n} \,,
\label{eqn:McalRelation}
\end{equation}
where $U_{x_1 \cdots x_n}$ is an ``evanescent term'' (see \cref{eqn:Ustructure} below for a detailed expression), which satisfies
\begin{equation}
\tpsi^{x_1}(\pon_1) \cdots \tpsi^{x_n}(\pon_n)\, \big( U_{x_1 \cdots x_n} |_{J=0} \big) = 0 \,,
\label{eqn:Uproperty}
\end{equation}
where $\pon_i$ is an on-shell momentum. Therefore, $U_{x_1 \cdots x_n}$ does not contribute to on-shell amplitudes. As a consequence, the on-shell amplitudes encoded in $\tGamma[\tphi]$ and $\Gamma[\phi]$ are the same: 
\begin{equation}
\tAmp\left(\pon_1, \cdots, \pon_n\right) = \Amp\left(\pon_1, \cdots, \pon_n\right) \,. \vspace{2mm}
\end{equation}
\end{minipage}
\end{center}
\end{tcolorbox}
\vspace{1mm}\noindent
This is the main result of this paper. We emphasize that this result holds to all loop orders, since \cref{eqn:GammaToAmp} holds to all loop orders.

In \cref{subsec:Applications}, we will apply the above statement to show that tree-level and one-loop amplitudes are invariant under general field redefinitions. Concretely, we will parameterize a general field redefinition by writing the old fields $\eta(x)$ and the new fields $\teta(x)$ as functionals of each other:
\begin{equation}
\eta \;\longrightarrow\; \teta \;: \qquad
\eta = f \big[ \teta \big] \,.
\label{eqn:GeneralFieldRedefAdv}
\end{equation}
This accommodates all field redefinitions that are expected to leave the $S$-matrix elements invariant, and in particular includes field redefinitions that involve derivatives. Such a field redefinition leads to a new Lagrangian, which gives a new 1PI effective action $\tGamma[\tphi]$. In \cref{subsec:Applications}, we will show that up to one-loop order, one can find an analytic functional relation $\phi[\tphi]$, such that the new 1PI effective action is related to the old one as in \cref{eqn:GammaRelation}, $\tGamma[\tphi] = \Gamma\big[\phi[\tphi]\big]$. Therefore, the transformation lemma applies, which leads to the conclusion that the on-shell amplitudes are the same.

\subsection{Proof of the Transformation Lemma}
\label{subsec:Proof}
Given a relation between two 1PI effective actions $\tGamma[\tphi]$ and $\Gamma[\phi]$ as in \cref{eqn:GammaRelation}:
\begin{equation}
\tGamma [\tphi] = \Gamma \big[ \phi[\tphi] \big] \,,
\label{eqn:GammaRelationRedo}
\end{equation}
we now address how their corresponding amplitudes would be related. Specifically, we will prove the transformation lemma described above; see \cref{eqn:GammaRelation,eqn:McalRelation,eqn:Uproperty}. Following the procedure in \cref{eqn:GammaToAmp}, we will first use \cref{eqn:GammaRelationRedo} to derive the relations between their functional derivatives, and then the relations between the amputated correlation functions $\tMcal_{x_1 \cdots x_n}$ and $\Mcal_{x_1 \cdots x_n}$, and eventually the relations between the amplitudes.

\subsubsection*{Zero Source Condition}

We begin by relating the first functional derivatives of the two effective actions. They are related by the chain rule
\begin{equation}
\frac{\delta\tGamma}{\delta\tphi^x} = \frac{\delta\phi^y}{\delta\tphi^x} \frac{\delta\Gamma}{\delta\phi^y} \,.
\label{eqn:del1GammaRelation}
\end{equation}
It means that for analytic functional relations $\phi[\tphi]$ in \cref{eqn:GammaRelationRedo}, where the matrix $\delta\phi^y/\delta\tphi^x$ is invertible, the zero source condition \cref{eqn:phiv} is unchanged:
\begin{equation}
\frac{\delta\tGamma}{\delta\tphi^x} \bigg|_{\tphi=\tphi_v} = 0
\qquad\Longleftrightarrow\qquad
\frac{\delta\Gamma}{\delta\phi^x} \bigg|_{\phi=\phi[ \tphi_v ]} = 0 \,.
\label{eq:TransformationZeroSourceCondition}
\end{equation}
Put in other words, $\phi_v(x)$ is given by plugging $\tphi_v(x)$ into the functional relation $\phi[\tphi]$:
\begin{equation}
\phi_v(x) = \phi \big[ \tphi_v \big](x) \,.
\label{eqn:phivRelation}
\end{equation}
Note that this would not be true if there were an inhomogeneous piece in \cref{eqn:del1GammaRelation}.

\subsubsection*{On-shell Condition}

Now we move onto the relation between the second derivatives. Following \cref{eqn:del1GammaRelation}, we derive the relation between the second functional derivatives again using the chain rule:
\begin{equation}
\frac{\delta^2 \tGamma}{\delta\tphi^{x_1} \delta\tphi^{x_2}}
= \frac{\delta\phi^{y_1}}{\delta\tphi^{x_1}} \frac{\delta\phi^{y_2}}{\delta\tphi^{x_2}}
\frac{\delta^2 \Gamma}{\delta\phi^{y_1} \delta\phi^{y_2}}
+ \frac{\delta^2 \phi^{y_1}}{\delta\tphi^{x_1} \delta\tphi^{x_2}} \frac{\delta\Gamma}{\delta\phi^{y_1}} \,.
\label{eqn:del2GammaRelation}
\end{equation}
From \cref{eq:TransformationZeroSourceCondition}, we see that the inhomogeneous piece vanishes when this expression is evaluated at $\tphi(x)=\tphi_v(x)$:
\begin{equation}
\frac{\delta^2 \tGamma}{\delta\tphi^{x_1} \delta\tphi^{x_2}} \bigg|_{\tphi_v}
= \left( \frac{\delta\phi^{y_1}}{\delta\tphi^{x_1}} \bigg|_{\tphi_v} \right)
\left( \frac{\delta\phi^{y_2}}{\delta\tphi^{x_2}} \bigg|_{\tphi_v} \right)
\left( \frac{\delta^2 \Gamma}{\delta\phi^{y_1} \delta\phi^{y_2}} \bigg|_{\phi_v} \right) \,,
\label{eqn:del2GammaRelationvev}
\end{equation}
where we have used \cref{eqn:phivRelation} for the last factor. This tells us that the on-shell momentum condition \cref{eqn:OnShellpGamma} is unchanged:
\begin{equation}
\int \dd^4x_1\, e^{i\pon x_1}\, \frac{\delta^2\Gamma}{\delta\phi^{x_1} \delta\phi^{x_2}} \bigg|_{\phi_v} = 0
\qquad\Longleftrightarrow\qquad
\int \dd^4x_1\, e^{i\pon x_1}\, \frac{\delta^2\tGamma}{\delta\tphi^{x_1} \delta\tphi^{x_2}} \bigg|_{\tphi_v} = 0 \,,
\end{equation}
again for analytic functional relations $\phi[\tphi]$ such that the matrix $\delta\phi^y/\delta\tphi^x$ is invertible. Moreover, from \cref{eqn:psixOnShellp} we see that \cref{eqn:del2GammaRelationvev} also implies the following relation between the \emph{on-shell} external wavefunctions
\begin{equation}
\psi^{y}(\pon) = \left( \frac{\delta\phi^y}{\delta\tphi^x} \bigg|_{\tphi_v} \right) \tpsi^x(\pon) \,.
\label{eqn:psixRelation}
\end{equation}
Note however that eigenstates with nonzero eigenvalues $\psi^y(p)$ and $\tpsi^x(p)$ with off-shell momentum $p^\mu$ are not related in such a simple way. This is because \cref{eqn:del2GammaRelationvev} is a congruence transform instead of a similarity transform between the two matrices $\frac{\delta^2 \tGamma}{\delta\tphi^{x_1} \delta\tphi^{x_2}} \big|_{\tphi_v}$ and $\frac{\delta^2 \Gamma}{\delta\phi^{y_1} \delta\phi^{y_2}} \big|_{\phi_v}$. Under such a transform, the nonzero eigenvalues are not preserved/invariant, which is also inferred by the mismatch regarding the upper/lower index structure between the two sides of \cref{eqn:psixEigen}.

\clearpage

\subsubsection*{Three-point Function}

Following \cref{eqn:del2GammaRelation}, one can further move on to the third functional derivatives of $\Gamma[\phi]$, which are of course the three-point amputated correlation functions (\cf\ \cref{eqn:M3}):
\begin{align}
\tMcal_{x_1 x_2 x_3} &=
\frac{\delta\phi^{y_1}}{\delta\tphi^{x_1}}
\frac{\delta\phi^{y_2}}{\delta\tphi^{x_2}}
\frac{\delta\phi^{y_3}}{\delta\tphi^{x_3}}\, \Mcal_{y_1 y_2 y_3}
- \frac{\delta^3 \phi^{y_1}}{\delta\tphi^{x_1} \delta\tphi^{x_2} \delta\tphi^{x_3}} \frac{\delta\Gamma}{\delta\phi^{y_1}}
\notag\\[5pt]
&\quad
- \left( \frac{\delta^2 \phi^{y_1}}{\delta\tphi^{x_2} \delta\tphi^{x_3}} \frac{\delta\phi^{y_2}}{\delta\tphi^{x_1}} + \frac{\delta^2 \phi^{y_1}}{\delta\tphi^{x_1} \delta\tphi^{x_3}} \frac{\delta\phi^{y_2}}{\delta\tphi^{x_2}} + \frac{\delta^2 \phi^{y_1}}{\delta\tphi^{x_1} \delta\tphi^{x_2}} \frac{\delta\phi^{y_2}}{\delta\tphi^{x_3}} \right) \frac{\delta^2\Gamma}{\delta\phi^{y_1} \delta\phi^{y_2}} \,.
\label{eqn:M3Relation}
\end{align}
We see that this expression involve more inhomogeneous pieces as compared to the second functional derivatives. However, these terms will drop out when computing on-shell amplitudes:
\begin{equation}
(2\pi)^4 \delta^4(\pon_1 + \pon_2 + \pon_3)\, i\tAmp \left( \pon_1, \pon_2, \pon_3 \right)
= \tpsi^{x_1}(\pon_1)\, \tpsi^{x_2}(\pon_2)\, \tpsi^{x_3}(\pon_3)\,
\Bigl( -i\tMcal_{x_1 x_2 x_3} \big|_{\tphi_v} \Bigr) \,.
\end{equation}
This is because the inhomogeneous pieces in the first and second lines of \cref{eqn:M3Relation} respectively  contain the following two types of factors:
\begin{subequations}\label{eqn:Ufactors}
\begin{align}
\frac{\delta\Gamma}{\delta\phi^y}:&\qquad
\frac{\delta\Gamma}{\delta\phi^y} \bigg|_{\phi_v} = 0 \,, \label{eqn:Ufactor1} \\[5pt]
\frac{\delta\phi^{y_2}}{\delta\tphi^{x_i}} \frac{\delta^2\Gamma}{\delta\phi^{y_1} \delta\phi^{y_2}}:&\qquad
\tpsi^{x_i}(\pon_i) \left( \frac{\delta\phi^{y_2}}{\delta\tphi^{x_i}} \bigg|_{\tphi_v} \right) \left( \frac{\delta^2\Gamma}{\delta\phi^{y_1} \delta\phi^{y_2}} \bigg|_{\phi_v} \right) = 0 \,, \label{eqn:Ufactor2}
\end{align}
\end{subequations}
where $x_i$ refers to an index in $\tMcal_{x_1 \cdots x_n}$, corresponding to an external leg of the diagram. As indicated above, terms with the first type of factors vanish upon enforcing the zero source condition $\tphi(x)=\tphi_v(x)$; terms with the second type of factors are nonzero at $\tphi(x)=\tphi_v(x)$, but will vanish upon a further contraction with the on-shell external wavefunctions $\tpsi^{x_i}(\pon_i)$, due to \cref{eqn:psixOnShellp,eqn:psixRelation}. Since they do not change the observable (on-shell) physics, we refer to these quantities as ``evanescent.'' A general parameterization of the evanescent terms that can appear is
\begin{equation}
U_{x_1 \cdots x_n} = a_{x_1 \cdots x_n}^{y_1}\, \frac{\delta\Gamma}{\delta\phi^{y_1}}
+ \sum_{i=1}^n b_{x_1 \cdots \hat{x}_i \cdots x_n}^{y_1}\, \frac{\delta\phi^{y_2}}{\delta\tphi^{x_i}} \frac{\delta^2\Gamma}{\delta\phi^{y_1} \delta\phi^{y_2}} \,.
\label{eqn:Ustructure}
\end{equation}
By construction, it satisfies the condition in \cref{eqn:Uproperty}.

\clearpage

Now we can rewrite \cref{eqn:M3Relation} as
\begin{equation}
\tMcal_{x_1 x_2 x_3} =
\frac{\delta\phi^{y_1}}{\delta\tphi^{x_1}}
\frac{\delta\phi^{y_2}}{\delta\tphi^{x_2}}
\frac{\delta\phi^{y_3}}{\delta\tphi^{x_3}}\, \Mcal_{y_1 y_2 y_3} + U_{x_1 x_2 x_3} \,,
\label{eqn:McalRelation3}
\end{equation}
where all the inhomogeneous terms are collectively denoted by the evanescent term $U_{x_1 x_2 x_3}$, which has the structure of \cref{eqn:Ustructure} (and hence satisfies the condition in \cref{eqn:Uproperty}). This proves the $n=3$ case of the transformation lemma around \cref{eqn:McalRelation,eqn:Uproperty}. It says that the three-point amputated correlation functions $\tMcal_{x_1 x_2 x_3}$ and $\Mcal_{y_1 y_2 y_3}$ are related homogeneously by the transformation matrices $\delta\phi^{y_i}/\delta\tphi^{x_i}$, up to an evanescent term $U_{x_1 x_2 x_3}$ that would not change the on-shell amplitudes $\tAmp\left(\pon_1, \pon_2, \pon_3\right)$.

\subsubsection*{$n$-point Functions}

The relation in \cref{eqn:McalRelation3} (with the structure of the evanescent term given in \cref{eqn:Ustructure}) holds also for higher-point amputated correlation functions, \ie, \cref{eqn:McalRelation}. To show this, one can derive the higher-point analog of \cref{eqn:M3Relation}, and then check if the inhomogeneous pieces are evanescent.

In order to organize the proof, we will use the recursive expression of the $n$-point functions
in \cref{eqn:M4Recursion,eqn:Recursion}, where an $(n+1)$-point amputated correlation function is concisely written in terms of the $n$-point ones. The $n=3$ case that we proved above in \cref{eqn:McalRelation3} (with \cref{eqn:Ustructure}) serves as the \emph{base case} for the induction. To further prove the result for arbitrary integer $n\ge 3$, we need to prove the \emph{induction step}: if \cref{eqn:McalRelation} (with \cref{eqn:Ustructure}) holds for $k$, then it will also hold for $k+1$.

To show this, we assume that $\tMcal_{x_1 \cdots x_k}$ and $\Mcal_{y_1 \cdots y_k}$ are related as in \cref{eqn:McalRelation}:
\begin{equation}
\tMcal_{x_1 \cdots x_k} = \frac{\delta\phi^{y_1}}{\delta\tphi^{x_1}} \cdots \frac{\delta\phi^{y_k}}{\delta\tphi^{x_k}}\, \Mcal_{y_1 \cdots y_k} + U_{x_1 \cdots x_k} \,,
\label{eqn:McalRelationk}
\end{equation}
where $U_{x_1 \cdots x_k}$ is an evanescent term that has the form in \cref{eqn:Ustructure}:
\begin{equation}
U_{x_1 \cdots x_k} = a_{x_1 \cdots x_k}^{y_1}\, \frac{\delta\Gamma}{\delta\phi^{y_1}}
+ \sum_{i=1}^k b_{x_1 \cdots \hat{x}_i \cdots x_k}^{y_1}\, \frac{\delta\phi^{y_2}}{\delta\tphi^{x_i}} \frac{\delta^2\Gamma}{\delta\phi^{y_1} \delta\phi^{y_2}} \,.
\label{eqn:Ustructurek}
\end{equation}

\clearpage

\noindent We then make use of the recursion relations for both $\tMcal$ and $\Mcal$
\begin{subequations}\label{eqn:ktokplus1}
\begin{align}
\tMcal_{x_1 \cdots x_k} &\quad\longrightarrow\quad
\tMcal_{x_1 \cdots x_k x_{k+1}} = \frac{\delta}{\delta\tphi^{x_{k+1}}}\, \tMcal_{x_1 \cdots x_k}
- \sum_{i=1}^k \tChris_{x_{k+1} x_i}^y\, \tMcal_{x_1 \cdots \hat{x}_i y \cdots x_k} \,, \\[5pt]
\Mcal_{y_1 \cdots y_k} &\quad\longrightarrow\quad
\Mcal_{y_1 \cdots y_k y_{k+1}} = \frac{\delta}{\delta\phi^{y_{k+1}}}\, \Mcal_{y_1 \cdots y_k}
- \sum_{i=1}^k \Chris_{y_{k+1} y_i}^z\, \Mcal_{y_1 \cdots \hat{y}_i z \cdots y_k} \,,
\end{align}
\end{subequations}
to show that consequently $\tMcal_{x_1 \cdots x_k x_{k+1}}$ and $\Mcal_{y_1 \cdots y_k y_{k+1}}$ will also be related as in \cref{eqn:McalRelation}. To this end, we compute the inhomogeneous pieces at $k+1$:
\begin{align}
&\tMcal_{x_1 \cdots x_k x_{k+1}} - \tfrac{\delta\phi^{y_1}}{\delta\tphi^{x_1}} \cdots \tfrac{\delta\phi^{y_k}}{\delta\tphi^{x_k}}\, \tfrac{\delta\phi^{y_{k+1}}}{\delta\tphi^{x_{k+1}}}\, \Mcal_{y_1 \cdots y_k y_{k+1}}
\notag\\[8pt]
&\hspace{20pt}
= - \sum_{i=1}^k
\left( \tfrac{\delta\phi^{y_1}}{\delta\tphi^{x_1}} \cdots \widehat{\tfrac{\delta\phi^{y_i}}{\delta\tphi^{x_i}}} \cdots \tfrac{\delta\phi^{y_k}}{\delta\tphi^{x_k}} \right)
\Mcal_{y_1 \cdots \hat{y}_i z \cdots y_k} \notag\\[4pt]
&\hspace{72pt}\times\left( \tfrac{\delta\phi^z}{\delta\tphi^y}\, \tChris_{x_{k+1}x_i}^y
- \tfrac{\delta\phi^{y_{k+1}}}{\delta\tphi^{x_{k+1}}} \tfrac{\delta\phi^{y_i}}{\delta\tphi^{x_i}}\, \Chris_{y_{k+1}y_i}^z
- \tfrac{\delta^2\phi^z}{\delta\tphi^{x_{k+1}} \delta\tphi^{x_i}} \right)
\notag\\[5pt]
&\hspace{35pt}
+ \left( \frac{\delta}{\delta\tphi^{x_{k+1}}}\, U_{x_1 \cdots x_k}
- \sum_{i=1}^k \tChris_{x_{k+1}x_i}^y\, U_{x_1 \cdots \hat{x}_i y \cdots x_k} \right) \,.
\label{eqn:Inhomokplus1}
\end{align}
To obtain this result, we have used \cref{eqn:McalRelationk,eqn:ktokplus1}. Our goal is to show that the right hand side has the general structure given in \cref{eqn:Ustructure}, so that it is an evanescent term. Let us check that this is true for the first and the second terms in turn.

To check the evanescence of the first term in \cref{eqn:Inhomokplus1}, we need to study the relation between $\tChris_{x_{k+1}x_i}^y$ and $\Chris_{y_{k+1}y_i}^z$. Recalling the definition in \cref{eqn:Chrisdef} and the relation in \cref{eqn:Propdef}, we get
\begin{equation}
\Chris_{x_1 x_2}^y = i\Mcal_{x_1 x_2 z}\, D^{zy}
= - \Mcal_{x_1 x_2 z} \left( \frac{\delta^2\Gamma}{\delta\phi^z \delta\phi^y} \right)^{-1} \,.
\label{eqn:ChrisMcalGamma}
\end{equation}
Therefore, using the relations in \cref{eqn:del2GammaRelation,eqn:M3Relation}, we have
\begin{align}
\tChris_{x_{k+1} x_i}^y &= - \tMcal_{x_{k+1} x_i u} \left( \tfrac{\delta^2\tGamma}{\delta\tphi^u \delta\tphi^y} \right)^{-1}
\notag\\[8pt]
&= - \left( \tfrac{\delta\phi^{y_{k+1}}}{\delta\tphi^{x_{k+1}}} \tfrac{\delta\phi^{y_i}}{\delta\tphi^{x_i}}
\tfrac{\delta\phi^v}{\delta\tphi^u}\, \Mcal_{y_{k+1} y_i v}
- \tfrac{\delta^2\phi^{y_1}}{\delta\tphi^{x_{k+1}} \delta\tphi^{x_i}}
\tfrac{\delta\phi^{y_2}}{\delta\tphi^u}
\tfrac{\delta^2 \Gamma}{\delta\phi^{y_1} \delta\phi^{y_2}} \right)
\tfrac{\delta\tphi^u}{\delta\phi^w} \left( \tfrac{\delta^2\Gamma}{\delta\phi^w \delta\phi^z} \right)^{-1} \tfrac{\delta\tphi^y}{\delta\phi^z}
%\notag\\[4pt]
%& \hspace{14pt}
+ U_{x_{k+1} x_i}^y
\notag\\[8pt]
&= \frac{\delta\phi^{y_{k+1}}}{\delta\tphi^{x_{k+1}}}
\frac{\delta\phi^{y_i}}{\delta\tphi^{x_i}}
\frac{\delta\tphi^y}{\delta\phi^z}\, \Chris_{y_{k+1} y_i}^z
+ \frac{\delta^2 \phi^z}{\delta\tphi^{x_{k+1}} \delta\tphi^{x_i}} \frac{\delta\tphi^y}{\delta\phi^z}
+ U_{x_{k+1} x_i}^y \,,
\label{eqn:ChrisRelation}
\end{align}
where $U_{x_{k+1} x_i}^y$ collects terms that contain the evanescent factors in \cref{eqn:Ufactors}, in a similar fashion as in \cref{eqn:Ustructure}. We emphasize that in the parentheses of the second line above, the second term is not evanescent and hence did not get collected into $U_{x_{k+1} x_i}^y$. This is because unlike $x_{k+1}$ or $x_i$, the index $u$ is not a leg, since it not an index in $\tMcal_{x_1 \cdots x_{k+1}}$. It yields the non-evanescent inhomogeneous piece in the last line. With the relation in \cref{eqn:ChrisRelation}, the first line of the result in \cref{eqn:Inhomokplus1} simplifies into
\begin{equation}
- \sum_{i=1}^k
\left( \frac{\delta\phi^{y_1}}{\delta\tphi^{x_1}} \cdots \widehat{\frac{\delta\phi^{y_i}}{\delta\tphi^{x_i}}} \cdots \frac{\delta\phi^{y_k}}{\delta\tphi^{x_k}} \right)
\Mcal_{y_1 \cdots \hat{y}_i z \cdots y_k}\, \frac{\delta\phi^z}{\delta\tphi^y}\, U_{x_{k+1} x_i}^y
\;\;\in\;\;
U_{x_1 \cdots x_{k+1}} \,.
\label{eqn:Line1Evan}
\end{equation}
As indicated here, this is clearly an evanescent term, because of the $U_{x_{k+1} x_i}^y$ factor.

Now let us move on to the second term in \cref{eqn:Inhomokplus1}. This term contains the evanescent term $U_{x_1 \cdots x_k}$, whose general form --- given in \cref{eqn:Ustructurek} --- comprises ``$a$-type'' and ``$b$-type'' evanescent factors in \cref{eqn:Ufactor1,eqn:Ufactor2}, respectively. However, if one takes a functional derivative $\frac{\delta}{\delta\tphi^{x_{k+1}}}$, and/or makes an index replacement $x_i \to y$, an evanescent term of the $a$-type or $b$-type might become non-evanescent. In what follows, we show that despite this, the combination in the second term of \cref{eqn:Inhomokplus1} is still evanescent.

Since the second term of \cref{eqn:Inhomokplus1} is linear in $U_{x_1 \cdots x_k}$, we can examine its $a$-type and $b$-type terms separately. Let us begin with the $a$-type terms. The evanescence of an $a$-type term does not rely on any of its indices being a leg $x_i$, so the index replacement $x_i \to y$ would not cause any problems. On the other hand, it does rely on containing a factor of the first functional derivative of $\Gamma$, so the additional functional derivative could potentially be a problem. However, since this additional functional derivative is at a leg $x_{k+1}$, any potentially problematic term that arises from an $a$-type term will simply be a $b$-type term, which is still evanescent:
\begin{equation}
\frac{\delta}{\delta\tphi^{x_{k+1}}} \left( a_{x_1 \cdots x_k}^{y_1}\, \frac{\delta\Gamma}{\delta\phi^{y_1}} \right) \supset a_{x_1 \cdots x_k}^{y_1} \frac{\delta\phi^{y_2}}{\delta\tphi^{x_{k+1}}} \frac{\delta^2 \Gamma}{\delta\phi^{y_1} \delta\phi^{y_2}}
\;\;\in\;\;
U_{x_1 \cdots x_{k+1}} \,.
\end{equation}
Therefore, for $a$-type terms in $U_{x_1 \cdots x_k}$, the second term in \cref{eqn:Inhomokplus1} remains evanescent for each individual term in its parentheses.

Now let us check the $b$-type terms. Acting the additional functional derivative on them yields the following non-evanescent terms
\begin{align}
&\frac{\delta}{\delta\tphi^{x_{k+1}}}\, U_{x_1 \cdots x_k} \supset
\frac{\delta}{\delta\tphi^{x_{k+1}}} \left( \sum_{i=1}^k b_{x_1 \cdots \hat{x}_i \cdots x_k}^{y_1}\, \frac{\delta\phi^{y_2}}{\delta\tphi^{x_i}} \frac{\delta^2\Gamma}{\delta\phi^{y_1} \delta\phi^{y_2}} \right)
\notag\\[5pt]
&\qquad
\supset
\sum_{i=1}^k b_{x_1 \cdots \hat{x}_i \cdots x_k}^{y_1} \left( \frac{\delta^2\phi^{y_2}}{\delta\tphi^{x_{k+1}} \delta\tphi^{x_i}} \frac{\delta^2\Gamma}{\delta\phi^{y_1} \delta\phi^{y_2}} + \frac{\delta\phi^{y_2}}{\delta\tphi^{x_i}} \frac{\delta\phi^{y_{k+1}}}{\delta\tphi^{x_{k+1}}} \frac{\delta^3\Gamma}{\delta\phi^{y_1} \delta\phi^{y_2} \delta\phi^{y_{k+1}}} \right) \,.
\label{eqn:btypenonU1}
\end{align}
On the other hand, the external-to-internal index replacement $x_i \to y$ yields the following non-evanescent terms
\begin{align}
&-\sum_{i=1}^k \tChris_{x_{k+1} x_i}^y U_{x_1 \cdots \hat{x}_i y \cdots x_k}
\supset - \sum_{i=1}^k \tChris_{x_{k+1} x_i}^y b_{x_1 \cdots \hat{x}_i \cdots x_k}^{y_1}\, \frac{\delta\phi^{y_2}}{\delta\tphi^y} \frac{\delta^2\Gamma}{\delta\phi^{y_1} \delta\phi^{y_2}}
\notag\\[5pt]
&\qquad
\supset -\sum_{i=1}^k \left( \frac{\delta\phi^{y_{k+1}}}{\delta\tphi^{x_{k+1}}} \frac{\delta\phi^{y_i}}{\delta\tphi^{x_i}} \frac{\delta\tphi^y}{\delta\phi^z}\, G_{y_{k+1} y_i}^z + \frac{\delta^2\phi^z}{\delta\tphi^{x_{k+1}} \delta\tphi^{x_i}} \frac{\delta\tphi^y}{\delta\phi^z} \right) b_{x_1 \cdots \hat{x}_i \cdots x_k}^{y_1}\, \frac{\delta\phi^{y_2}}{\delta\tphi^y} \frac{\delta^2\Gamma}{\delta\phi^{y_1} \delta\phi^{y_2}}
\notag\\[5pt]
&\qquad
\supset -\sum_{i=1}^k b_{x_1 \cdots \hat{x}_i \cdots x_k}^{y_1} \left( 
\frac{\delta\phi^{y_{k+1}}}{\delta\tphi^{x_{k+1}}}
\frac{\delta\phi^{y_i}}{\delta\tphi^{x_i}}\, G_{y_{k+1} y_i}^{y_2}
+ \frac{\delta^2\phi^{y_2}}{\delta\tphi^{x_{k+1}} \delta\tphi^{x_i}} \right)
\frac{\delta^2\Gamma}{\delta\phi^{y_1} \delta\phi^{y_2}}
\notag\\[5pt]
&\qquad
\supset -\sum_{i=1}^k b_{x_1 \cdots \hat{x}_i \cdots x_k}^{y_1} \left(
\frac{\delta\phi^{y_{k+1}}}{\delta\tphi^{x_{k+1}}}
\frac{\delta\phi^{y_i}}{\delta\tphi^{x_i}}
\frac{\delta^3\Gamma}{\delta\phi^{y_{k+1}} \delta\phi^{y_i} \delta\phi^{y_1}}
+ \frac{\delta^2\phi^{y_2}}{\delta\tphi^{x_{k+1}} \delta\tphi^{x_i}} \frac{\delta^2\Gamma}{\delta\phi^{y_1} \delta\phi^{y_2}} \right) \,,
\label{eqn:btypenonU2}
\end{align}
where we have used the results in \cref{eqn:ChrisRelation}, \cref{eqn:ChrisMcalGamma}, and then \cref{eqn:M3}. We see that the non-evanescent terms in \cref{eqn:btypenonU2} precisely cancel those from \cref{eqn:btypenonU1}. Therefore, for $b$-type terms in $U_{x_1 \cdots x_k}$, the second term in \cref{eqn:Inhomokplus1} remains evanescent as a sum of the two terms in its parentheses.

Combining our investigations on $a$-type and $b$-type terms in $U_{x_1 \cdots x_k}$, we conclude that the second term of the result in \cref{eqn:Inhomokplus1} remains evanescent:
\begin{equation}
\frac{\delta}{\delta\tphi^{x_{k+1}}}\, U_{x_1 \cdots x_k}
- \sum_{i=1}^k \tChris_{x_{k+1}x_i}^y\, U_{x_1 \cdots \hat{x}_i y \cdots x_k}
\;\;\in\;\;
U_{x_1 \cdots x_{k+1}} \,.
\label{eqn:Line2Evan}
\end{equation}
\cref{eqn:Line1Evan,eqn:Line2Evan} together then complete our proof of the induction step, namely that the following relation for $(k+1)$-point amputated correlation functions holds
\begin{equation}
\tMcal_{x_1 \cdots x_k x_{k+1}} =
\frac{\delta\phi^{y_1}}{\delta\tphi^{x_1}} \cdots \frac{\delta\phi^{y_k}}{\delta\tphi^{x_k}}\, \frac{\delta\phi^{y_{k+1}}}{\delta\tphi^{x_{k+1}}}\, \Mcal_{y_1 \cdots y_k y_{k+1}}
+ U_{x_1 \cdots x_k x_{k+1}} \,,
\end{equation}
provided that it holds for $k$-point functions (\cref{eqn:McalRelationk}). Combining this induction step with the base case that we proved for $n=3$ in \cref{eqn:McalRelation3}, this proves that \cref{eqn:McalRelation} (together with \cref{eqn:Ustructure}) holds for an arbitrary integer $n\ge 3$.

To complete our proof of the transformation lemma, let us show that \cref{eqn:McalRelation,eqn:Ustructure} imply that the on-shell amplitudes are the same:
\begin{align}
&(2\pi)^4 \delta^4(\pon_1 +\cdots + \pon_n)\, i\tAmp \left( \pon_1, \cdots, \pon_n \right)
\notag\\[8pt]
&\hspace{40pt}
= \Big[ \tpsi^{x_1}(\pon_1) \cdots \tpsi^{x_n}(\pon_n) \Big]\,
\Bigl( -i\tMcal_{x_1 \cdots x_n}|_{\tphi_v} \Bigr)
\notag\\[5pt]
&\hspace{40pt}
= \Big[ \tpsi^{x_1}(\pon_1) \cdots \tpsi^{x_n}(\pon_n) \Big]\,
\left( \frac{\delta\phi^{y_1}}{\delta\tphi^{x_1}} \bigg|_{\tphi_v} \right) \cdots 
\left( \frac{\delta\phi^{y_n}}{\delta\tphi^{x_n}} \bigg|_{\tphi_v} \right)
\Bigl( -i\Mcal_{y_1 \cdots y_n}|_{\phi_v} \Bigr)
\notag\\[5pt]
&\hspace{40pt}
= \Big[ \psi^{y_1}(\pon_1) \cdots \psi^{y_n}(\pon_n) \Big]\,
\Bigl( -i\Mcal_{y_1 \cdots y_n}|_{\phi_v} \Bigr)
\notag\\[8pt]
&\hspace{40pt}
= (2\pi)^4 \delta^4(\pon_1 +\cdots + \pon_n)\, i\Amp \left( \pon_1, \cdots, \pon_n \right) \,,
\label{eqn:AmpInvariance}
\end{align}
where we have used the relation in \cref{eqn:psixRelation}. $\blacksquare$

\subsection{Applications to Tree and One-Loop Amplitudes}
\label{subsec:Applications}

We now apply the transformation lemma to show that tree-level and one-loop amplitudes are invariant under a general field redefinition that accommodates derivatives:
\begin{equation}
\eta \;\longrightarrow\; \teta \;: \qquad
\eta = f \big[ \teta \big] \,.
\label{eqn:GeneralFieldRedef}
\end{equation}
Our task is to show that under such a field redefinition, the change of the 1PI effective action can be described by \cref{eqn:GammaRelation}, \ie, $\tGamma [\tphi] = \Gamma \big[ \phi[\tphi] \big]$ for some $\phi[\tphi]$.

\subsubsection*{Tree-Level Amplitudes}

We begin with the tree-level case. When we perform a field redefinition described by \cref{eqn:GeneralFieldRedef}, the new action at tree level is simply given by substituting in that relation (see \cref{eqn:Stilde} for a general expression):
\begin{equation}
\tilde S \big[ \teta \big] = S \Big[ f \big[ \teta \big] \Big] \,.
\label{eqn:STransTree}
\end{equation}
Meanwhile, the tree-level 1PI effective action is just given by the action of the theory
\begin{equation}
\Gamma [\phi] = S [\phi] \,,
\qquad\text{and}\qquad
\tGamma [\tphi] = \tilde{S} [\tphi] \,.
\label{eqn:GammaTree}
\end{equation}
Putting \cref{eqn:STransTree,eqn:GammaTree} together, we get the relation between the new and old 1PI effective actions at the tree level:
\begin{equation}
\tGamma \big[ \tphi \big] = \tilde S \big[ \tphi \big]
= S \Big[ f \big[ \tphi \big] \Big] = \Gamma \Big[ f \big[ \tphi \big] \Big] \,.
\label{eqn:GammaTransTree}
\end{equation}
We see that they do satisfy the transformation relation in \cref{eqn:GammaRelation}, with the functional $\phi[\tphi]$ identified to be the field redefinition functional itself:
\begin{equation}
\phi[\tphi] = f[\tphi] \,.
\label{eqn:phitphiTree}
\end{equation}
The transformation lemma then implies that tree-level on-shell amplitudes are invariant under the general field redefinition in \cref{eqn:GeneralFieldRedef}.

\subsubsection*{One-Loop Amplitudes}

In the one-loop case, both \cref{eqn:STransTree,eqn:GammaTree} become more complicated. As elaborated in \cref{appsec:Invariance}, under a general field redefinition in \cref{eqn:GeneralFieldRedef}, the action in terms of the new field at the loop level is (see \cref{eqn:Stilde})
\begin{equation}
\tilde S \big[ \tilde\eta \big] = S \Big[ f \big[ \tilde\eta \big] \Big] - i\log\det \left( \frac{\delta\eta^y}{\delta\tilde\eta^x} \right) \,.
\label{eqn:STransLoop}
\end{equation}
Note the extra one-loop sized Jacobian term compared to \cref{eqn:STransTree}. Apart from anomalous fermion chiral transformations, this term vanishes if one works with dimensional regularization \cite{Arzt:1993gz, Manohar:2018aog}. However, we keep it here to make our argument independent of the choice of the regularization scheme. On the other hand, up to one-loop level, the relation between $\Gamma[\phi]$ and $S[\phi]$ is also modified:
\begin{equation}
\Gamma [\phi] = S [\phi] + \frac{i}{2} \log\det \left( \frac{\delta^2 S}{\delta\eta^{x_1} \delta\eta^{x_2}} \bigg|_{\eta=\phi} \right) \,.
\label{eqn:GammaLoop}
\end{equation}

With \cref{eqn:STransLoop,eqn:GammaLoop}, we can obtain the relation between $\tGamma[\tphi]$ and $\Gamma[\phi]$. The second functional derivatives of the new action are
\begin{equation}
\frac{\delta^2 \tilde{S}}{\delta\teta^{x_1} \delta\teta^{x_2}} = \frac{\delta\eta^{y_1}}{\delta\teta^{x_1}} \frac{\delta\eta^{y_2}}{\delta\teta^{x_2}} \frac{\delta^2 S}{\delta\eta^{y_1} \delta\eta^{y_2}}
+ \frac{\delta^2\eta^{y_1}}{\delta\teta^{x_1} \delta\teta^{x_2}} \frac{\delta S}{\delta \eta^{y_1}}
+ \left( \text{one-loop terms} \right) \,.
\end{equation}

\clearpage

\noindent Plugging this in, we obtain the following relation up to one-loop level
\begin{align}
\tGamma \big[ \tphi \big] &= \tilde{S} \big[ \tphi \big] + \frac{i}{2} \log\det \left( \frac{\delta^2 \tilde{S}}{\delta\teta^{x_1} \delta\teta^{x_2}} \bigg|_{\tphi} \right)
\notag\\[8pt]
&= S \Big[ f \big[ \tphi \big] \Big]
- i\log\det \left( \frac{\delta\eta^y}{\delta\teta^x} \bigg|_{\tphi} \right)
\notag\\[3pt]
&\qquad
+ \frac{i}{2} \log\det \left(
\frac{\delta\eta^{y_1}}{\delta\teta^{x_1}}
\frac{\delta\eta^{y_2}}{\delta\teta^{x_2}} \bigg|_{\tphi} \;
\frac{\delta^2 S}{\delta\eta^{y_1} \delta\eta^{y_2}} \bigg|_{f[\tphi]}
+ \frac{\delta^2\eta^{y_1}}{\delta\teta^{x_1} \delta\teta^{x_2}} \bigg|_{\tphi} \;
\frac{\delta S}{\delta \eta^{y_1}} \bigg|_{f[\tphi]}
\right)
\notag\\[8pt]
&= S \Big[ f \big[ \tphi \big] \Big]
+ \frac{i}{2} \log\det \left(
\frac{\delta^2 S}{\delta\eta^{y_1} \delta\eta^{y_2}} \bigg|_{f[\tphi]}
+ \frac{\delta^2\eta^y}{\delta\teta^{x_1} \delta\teta^{x_2}} \bigg|_{\tphi} \;
\frac{\delta\teta^{x_1}}{\delta\eta^{y_1}}
\frac{\delta\teta^{x_2}}{\delta\eta^{y_2}}
\frac{\delta S}{\delta \eta^y} \bigg|_{f[\tphi]}
\right)
\notag\\[8pt]
&= \Gamma \Big[ f \big[ \tphi \big] \Big] + \frac{i}{2} \Tr\log \left[ 1 +
\frac{\delta^2\eta^y}{\delta\teta^{x_1} \delta\teta^{x_2}} \bigg|_{\tphi} \;
\Big( \frac{\delta^2 S}{\delta\eta^{y_1} \delta\eta^{z_1}} \Big)^{-1}
\frac{\delta\teta^{x_1}}{\delta\eta^{y_1}}
\frac{\delta\teta^{x_2}}{\delta\eta^{y_2}}
\frac{\delta S}{\delta \eta^y} \bigg|_{f[\tphi]}
\right] \,.
\label{eqn:GammaTransLoop}
\end{align}
We see that there is an extra term compared to \cref{eqn:GammaTransTree}. Although this term looks complicated, after expanding the $\log$ it will yield a series of terms that are proportional to the tree-level equation of motion $\delta S/\delta\eta^y$, with some one-loop order coefficients $a^y$:
\begin{equation}
\tGamma \big[ \tphi \big] = \Gamma \Big[ f\big[\tphi\big] \Big] + a^y\big[\tphi\big] \left( \frac{\delta S}{\delta\eta^y} \bigg|_{f[\tphi]} \right) \,.
\end{equation}
Since $a^y[\tphi]$ are one-loop order, we can replace $S$ with $\Gamma$, as the difference introduced will be two-loop order. For the same reason, we can keep the accuracy only to the first power of $a^y[\tphi]$. Carrying out these manipulations, we get
\begin{equation}
\tGamma \big[ \tphi \big] = \Gamma \Big[ f\big[\tphi\big] \Big] + a^y\big[\tphi\big] \left( \frac{\delta \Gamma}{\delta\phi^y} \bigg|_{f[\tphi]} \right) = \Gamma \Big[ f\big[\tphi\big] + a\big[\tphi\big] \Big] \,.
\end{equation}
This shows that up to one-loop order, the 1PI effective actions $\tGamma [\tphi]$ and $\Gamma [\phi]$ again satisfy the transformation relation in \cref{eqn:GammaRelation}, where the functional $\phi[\tphi]$ is identified as
\begin{equation}
\phi[\tphi] = f\big[\tphi\big] + a\big[\tphi\big] \,.
\label{eqn:phitphiLoop}
\end{equation}
The transformation lemma then implies that on-shell amplitudes up to one-loop order are invariant under the field redefinition in \cref{eqn:GeneralFieldRedef}.

%%%%%%%%%%%%%%%%%%%%%%%%%%%%%%%%%%%%%%%%%%%%%%%%%%%%%%%%%%%%%%%%%%%%%%%%%%%%%%%%
\section{Towards a Geometric Interpretation}
\label{sec:Geometry}
%%%%%%%%%%%%%%%%%%%%%%%%%%%%%%%%%%%%%%%%%%%%%%%%%%%%%%%%%%%%%%%%%%%%%%%%%%%%%%%%

In the previous section, we used the transformation lemma to show that the tree-level and one-loop amputated correlation functions transform as tensors under generalized field redefinitions up to terms that vanish when the sources are set to zero and the external states are taken to be on shell. Since the proof of the transformation lemma in \cref{subsec:Proof} is rather technical, it would be useful to have a more intuitive explanation of these results. With this motivation in mind, we note that the recursion relation \cref{eqn:Recursion} appears to have a tensor-like structure --- the relation between the $n+1$ and the $n$ leg results resembles that of a covariant derivative acting on a tensor:
\begin{equation}
\Mcal_{x_1 \cdots x_n x_{n+1}} = \frac{\delta}{\delta\phi^{x_{n+1}}}\, \Mcal_{x_1 \cdots x_n}
- \sum_{i=1}^n \Chris_{x_{n+1} x_i}^y\, \Mcal_{x_1 \cdots \hat{x}_i y \cdots x_n}
\quad\stackrel{?}{=}\quad
\nabla_{x_{n+1}} \Mcal_{x_1 \cdots x_n} \,.
\label{eqn:RecursionGeo}
\end{equation}
Having the notion of a covariant derivative evokes the expectation that this can be used to define parallel transport along some kind of geometric space.

However, such a geometric interpretation requires that we can identify a manifold such that
\begin{subequations}\label{eqn:GeoConditions}
\begin{align}
&\!\!\!\text{a)\,\,The functional derivative $\frac{\delta}{\delta\phi^x}$ can be interpreted as a coordinate derivative.}
\label{eqn:GeoCondition1} \\[5pt]
&\!\!\!\text{b)\,\,The factor $\Chris_{x_1 x_2}^y$ serves as a connection.}
\label{eqn:GeoCondition2} \\[8pt]
&\!\!\!\text{c)\,\,The amputated correlation functions $\Mcal_{x_1 \cdots x_n}$ transform as tensors.}
\label{eqn:GeoCondition3}
\end{align}
\end{subequations}
We already know that the third condition fails. When the transformation lemma applies, the amputated correlation functions do not transform as tensors, due to the extra evanescent term:
\begin{equation}
\tGamma [\tphi] = \Gamma \big[ \phi[\tphi] \big]
\qquad\Longrightarrow\qquad
\tMcal_{x_1 \cdots x_n} = \frac{\delta\phi^{y_1}}{\delta\tphi^{x_1}} \cdots \frac{\delta\phi^{y_n}}{\delta\tphi^{x_n}}\, \Mcal_{y_1 \cdots y_n} + U_{x_1 \cdots x_n} \,.
\label{eqn:TransStatementGeo}
\end{equation}
Nevertheless, one could imagine that some sort of procedure to quotient out the evanescent terms exists, leaving behind a well defined ``projective'' geometry, that we will refer to as ``functional geometry.'' This section is devoted to exploring the possibility that the resulting ``functional manifold'' could be constructed. In particular, we will discuss aspects where this approach appears to be successful, and we will highlight some ways in which it fails. We will also comment on the relation between functional geometry and the well-established field space geometry formalism (which does not incorporate derivative field redefinitions).

\subsection{Evidence for a Functional Manifold: Success and Failure}
\label{subsec:FunctionalManifold}

We begin by checking the condition in \cref{eqn:GeoCondition1}. Our goal is to find a manifold on which the functional derivatives $\frac{\delta}{\delta\phi(x)}$ can be identified with coordinate derivatives. To this end, we consider the so-called ``field configuration space,'' which is the collection of all the $\phi(x)$ field configurations that are integrated over when computing the path integral. This space is naturally endowed with a functional differentiable structure, and hence can be viewed as a differential manifold, albeit an infinite dimensional one \cite{DeWitt:1984sjp, DeWitt:2003pm, DeWitt:2007mi}. We refer to this manifold as the ``functional manifold.''

One way to parameterize the field configuration space is to simply specify the values of the field at all the spacetime points:
\begin{equation}
\Big\{ \phi^x   \;\big|\;   x \in \text{spacetime} \Big\} \,.
\label{eqn:ConfigurationSpace}
\end{equation}
Each allowed value of the set of variables in \cref{eqn:ConfigurationSpace} gives a specific field configuration $\phi(x)$, and by our construction, corresponds to a specific point on the functional manifold. The whole functional manifold is a collection of all such points. The functional manifold is therefore charted by $\{\phi^x\}$. Functions on this manifold are functions of the field configurations, or equivalently functionals of the field $\phi^x$, for example the 1PI effective action $\Gamma[\phi]$. Therefore, functional derivatives with respect to the field $\phi^x$ are just coordinate derivatives on this manifold, and they form a basis for the tangent space:
\begin{equation}
\left\{ \frac{\delta}{\delta\phi^x} \quad\text{with}\;\; \phi^x \;\;\text{parameterizing the configuration space} \right\} \,.
\end{equation}

\subsubsection{Success: 1PI Effective Action as a Scalar}
\label{subsubsec:GammaScalar}

We argued in \cref{subsec:Applications} that under a general field redefinition parameterized by \cref{eqn:GeneralFieldRedefAdv}, the 1PI effective action (up to one-loop level) transforms as in \cref{eqn:TransStatementGeo}:
\begin{equation}
\tGamma [\tphi] = \Gamma \big[ \phi[\tphi] \big] \,.
\label{eqn:GammaRelationGeo}
\end{equation}
As a functional relation, $\phi[\tphi]$ is a map between the two field configurations, $\{\phi^x\}$ and $\{\tphi^x\}$. Alternatively, one can view this map as reparameterizing a point on the functional manifold $\{\phi^x\}$ to the same point using the new set of variables $\{\tphi^x\}$. Therefore, it is a re-charting or coordinate change on the functional manifold.\footnote{Note that this coordinate change $\phi[\tphi]$ is not necessarily the same as the field redefinition relation $\eta=f[\teta]$, \cf\ \cref{eqn:phitphiTree,eqn:phitphiLoop}.}
From this point of view, \cref{eqn:GammaRelationGeo} means that the 1PI effective action transforms as a scalar on the functional manifold.

\subsubsection{Success: Physical Vacuum as a Geometric Point}
\label{subsubsec:VacuumPoint}

Since the 1PI effective action transforms as a scalar, its first functional derivative transforms as a vector on the functional manifold (\cf\ \cref{eqn:del1GammaRelation}):
\begin{equation}
\frac{\delta\tGamma}{\delta\tphi^x} \bigg|_{\tphi}
= \frac{\delta\phi^y}{\delta\tphi^x} \bigg|_{\tphi} \;
\frac{\delta\Gamma}{\delta\phi^y} \bigg|_{\phi[\tphi]} \,.
\label{eqn:del1GammaRelationGeo}
\end{equation}
Recall from \cref{eqn:phiv} that the physical vacuum field configuration $\phi_v(x)$ (for the original theory $S[\eta]$) is determined by
\begin{equation}
\frac{\delta\Gamma}{\delta\phi^x} \bigg|_{\phi=\phi_v} = 0 \,.
\end{equation}
The transformation law in \cref{eqn:del1GammaRelationGeo} then implies that the physical vacuum is a geometric point on the functional manifold --- it is independent of the chart chosen, and its coordinates changing accordingly:
\begin{equation}
\phi_v(x) = \phi \big[ \tphi_v \big](x) \,.
\label{eqn:phivRelationGeo}
\end{equation}

\subsubsection{Failure: Evanescent Terms Ruin Covariance}
\label{subsubsec:Evanescence}

Given these successes, we move on to check the conditions in \cref{eqn:GeoCondition2,eqn:GeoCondition3}. Unfortunately, it turns out that the functional manifold considered above fails to satisfy these conditions. However, it is still enlightening to see how it fails, since this can provide guidance for alternative constructions.

First, we can check the properties of $\Chris_{x_1 x_2}^y$. We will argue that it does not have the appropriate transformation rules to be interpreted as a connection. Following standard methodology, we use $\{ \frac{\delta}{\delta\phi^x}, \delta\phi^x \}$ as the bases of the tangent and cotangent spaces of the functional manifold. Then a connection $\Gamma$ can be defined using
\begin{equation}
\delta \left( \frac{\delta}{\delta\phi^{y_3}} \right) \equiv \delta\phi^{y_2}\, \nabla_{\phi^{y_2}} \left( \frac{\delta}{\delta\phi^{y_3}} \right) \equiv \delta\phi^{y_2}\, \Gamma_{y_2 y_3}^{y_1} \left( \frac{\delta}{\delta\phi^{y_1}} \right) \,,
\label{eqn:FunctionalConnection}
\end{equation}
where $\Gamma_{y_2 y_3}^{y_1}$ are components of the connection (not to be confused with the 1PI effective action). Now consider a coordinate change $\phi[\tphi]$. The bases transform as tensors
\begin{subequations}\label{eqn:BasesTrans}
\begin{align}
\frac{\delta}{\delta\tphi^x} &= \frac{\delta\phi^y}{\delta\tphi^x}\, \frac{\delta}{\delta\phi^y} \,, \\[5pt]
\delta\tphi^x &= \frac{\delta\tphi^x}{\delta\phi^y}\, \delta\phi^y \,.
\end{align}
\end{subequations}
This leads to the following standard transformation law for a connection:
\begin{equation}
\tGamma_{x_2 x_3}^{x_1} = \frac{\delta\phi^{y_2}}{\delta\tphi^{x_2}} \frac{\delta\phi^{y_3}}{\delta\tphi^{x_3}}
\frac{\delta\tphi^{x_1}}{\delta\phi^{y_1}}\, \Gamma_{y_2 y_3}^{y_1}
+ \frac{\delta^2\phi^{y_1}}{\delta\tphi^{x_2} \delta\tphi^{x_3}} \frac{\delta\tphi^{x_1}}{\delta\phi^{y_1}} \,.
\label{eqn:ConnectionTrans}
\end{equation}

Now comparing with the transformation property of $\Chris_{x_1 x_2}^y$ derived in \cref{eqn:ChrisRelation}, we see that $\Chris_{x_1 x_2}^y$ does not satisfy \cref{eqn:ConnectionTrans}. Therefore, it cannot serve as a connection on the functional manifold. However, it is worth mentioning that the transformation property of $\Chris_{x_1 x_2}^y$ in \cref{eqn:ChrisRelation} is very close to that in \cref{eqn:ConnectionTrans}; the only difference is that \cref{eqn:ChrisRelation} contains an extra evanescent term. Given that $\Chris_{x_{n+1} x_i}^y$ does not serve as a connection on the functional manifold, the right-hand side of the recursion relation in \cref{eqn:RecursionGeo} cannot be interpreted as a covariant derivative ``$\nabla_{x_{n+1}}$''.

The same essential obstruction holds for the amputated correlation functions. As derived in \cref{sec:Invariance}, the transformation property of the amputated correlation functions are given in \cref{eqn:TransStatementGeo}. Clearly, they do not transform as tensors on the functional manifold, again due to the extra evanescent term. So similar to the situation of \cref{eqn:GeoCondition2}, the condition in \cref{eqn:GeoCondition3} is almost satisfied, except for the evanescent term.

\subsubsection{Failure: Vanishing Curvature Tensor}
\label{subsubsec:ZeroCurvature}

We will now identify another fundamental issue with the functional geometry picture as defined above. We show that if we ignore the evanescent term issue discussed above and mindlessly use $\Chris_{x_1 x_2}^y$ defined in \cref{eqn:Chrisdef} as a connection to compute the Riemann curvature tensor, then it vanishes. One straightforward way to see this follows directly from the recursion relation \cref{eqn:RecursionGeo} --- using it twice, we find
\begin{equation}
\Mcal_{x_1 \cdots x_n y z} = \nabla_z \nabla_y\, \Mcal_{x_1 \cdots x_n} \,.
\end{equation}
Then the crossing symmetry of $\Mcal_{x_1 \cdots x_n y z}$ between the legs $y$ and $z$ implies that
\begin{equation}
\comm{\nabla_y}{\nabla_z}\, \Mcal_{x_1 \cdots x_n} = 0 \,,
\end{equation}
namely that there is no curvature.  We will provide a bit more insight into this issue in \cref{sec:VanishingCurvatureFieldSpaceGeo} below, in terms of so-called field space geometry.

\subsection{Relation to the Field Space Geometry}
\label{subsec:FieldSpaceGeo}

There is a well-established geometric picture for amplitudes in the literature \cite{Honerkamp:1971sh, Tataru:1975ys, Alvarez-Gaume:1981exa, Alvarez-Gaume:1981exv, Vilkovisky:1984st, DeWitt:1984sjp, Gaillard:1985uh, DeWitt:1985sg, Alonso:2015fsp, Alonso:2016btr, Alonso:2016oah}, based on the idea of the ``field space manifold,'' which accommodates a narrower set of field redefinitions, namely those that do not involve derivatives. In this section, we comment on the relation between the functional manifold and the field space manifold. We will also discuss how a variety of quantities on the functional manifold reproduce geometric statements that have been derived using the field space geometry picture. Some of these have been shown in \cite{Cohen:2022uuw}. Here we give a more detailed discussion.

\subsubsection{Review of Field Space Geometry}
\label{subsubsec:ReviewFieldSpace}

We briefly review the field space geometry picture. For this purpose, we again focus on the case of scalar fields, similar with \cref{subsec:Correlation}. We consider an EFT of scalar fields $\{\phi^a\}$. The most general Lagrangian involving up to two derivatives is:
\begin{equation}
\Lag = - V (\phi) + \frac12\, g_{ab}(\phi) \big( \partial_\mu \phi^a \big) \big( \partial^\mu \phi^b \big) + \ord \big(\partial^4\big) \,.
\label{eqn:LagAJM}
\end{equation}
$V(\phi)$ and $g_{ab}(\phi)$ can be interpreted as functions on the so-called ``field space manifold,'' which consists of all the allowed field space (or target space) points. Note that each point on the field space manifold is specified by the set of values $\{\phi^a\}$, so it is a finite dimensional manifold, with its dimension being the number of field flavors. The field space geometry deals with the differential geometry on this manifold.

A field redefinition without derivatives
\begin{equation}
\phi = f \big( \tphi \big) \,,
\label{eqn:FieldRedefNod}
\end{equation}
can be viewed as a coordinate change on the field space manifold. As usual, the bases of its tangent and cotangent spaces $\{ \frac{\partial}{\partial\phi^a}, \dd\phi^a \}$ transform as tensors
\begin{subequations}
\begin{align}
\frac{\partial}{\partial\tphi^a} &= \frac{\partial\phi^b}{\partial\tphi^a}\, \frac{\partial}{\partial\phi^b} \,, \\[5pt]
\dd\tphi^a &= \frac{\partial\tphi^a}{\partial\phi^b}\, \dd\phi^b \,.
\end{align}
\end{subequations}
Using these bases, a connection on the manifold can be introduced as
\begin{equation}
\dd \left( \frac{\partial}{\partial\phi^c} \right) \equiv \dd\phi^b\, \nabla_{\phi^b} \left( \frac{\partial}{\partial\phi^c} \right) \equiv \dd\phi^b\, \Gamma_{bc}^a \left( \frac{\partial}{\partial\phi^a} \right) \,,
\label{eqn:FieldSpaceConnection}
\end{equation}
where $\Gamma_{bc}^a$ are the connection components (not to be confused with 1PI effective actions).
A covariant derivative of a general tensor is then given by
\begin{equation}
\nabla_c\, T^{a\cdots}{}_{b\cdots} = \partial_c\, T^{a\cdots}{}_{b\cdots}
+ \Big( \Gamma_{ck}^a\, T^{k\cdots}{}_{b\cdots} + \cdots \Big)
- \Big( \Gamma_{cb}^k\, T^{a\cdots}{}_{k\cdots} + \cdots \Big) \,.
\label{eqn:nabladef}
\end{equation}

We note that the function $g_{ab}(\phi)$ transforms as a $(0, 2)$-tensor under the non-derivative field redefinition in \cref{eqn:FieldRedefNod}:
\begin{equation}
\tilde{g}_{ab} \big(\tphi\big) = \frac{\partial\phi^c}{\partial\tphi^a} \frac{\partial\phi^d}{\partial\tphi^b}\, g_{cd}(\phi) \,.
\end{equation}
This object is a natural choice of a metric on the field space manifold. If we require the connection in \cref{eqn:FieldSpaceConnection} to be compatible with this metric, \ie, $\nabla_c\, g_{ab}=0$, we get the usual Levi-Civita connection:
\begin{equation}
\Gamma_{bc}^a = \frac12\, g^{ak} \left( g_{kb,c} + g_{kc,b} - g_{bc,k} \right) \,,
\label{eqn:LeviCivita}
\end{equation}
where indices following a comma denote partial derivatives.

The field space geometry is a Riemannian geometry. On-shell amplitudes can be written in terms of geometric tensors on the field space manifold, multiplied by additional kinematic factors. For example, for the theory up to two-derivative interactions given in \cref{eqn:LagAJM}, the three-point amplitudes can be written as
\begin{equation}
- \left( \prod_{i=1}^3 \overline{g}_{a_i a_i}^{1/2} \right) \Amp_{\, a_1 a_2 a_3} \left( \pon_1, \pon_2, \pon_3 \right) = \overline{V}_{;\, (a_1 a_2 a_3)} \,.
\label{eqn:Amp3AJM}
\end{equation}
Here indices following a semicolon denote covariant derivatives under the Levi-Civita connection in \cref{eqn:LeviCivita}, and the parentheses denote a normalized symmetrization of these indices. The bars on the geometric quantities, $g_{ab}$, $V$, \etc indicates evaluating them at the physical vacuum point on the field space manifold. The four-point amplitudes have a similar but richer expression:
\begin{align}
- \left( \prod_{i=1}^4 \overline{g}_{a_i a_i}^{1/2} \right) \Amp_{\, a_1 a_2 a_3 a_4} \left( \pon_1, \pon_2, \pon_3, \pon_4 \right) &= \overline{V}_{;\, (a_1 a_2 a_3 a_4)} + \frac13\, \sum_{i<j} s_{ij} \overline{R}_{a_i (a_k a_l) a_j}
\notag\\[5pt]
&\quad
+ \left[ \overline{V}_{;\, (a_1 a_2 b)}\, \frac{\overline{g}^{bc}}{s_{12}-m_b^2}\, \overline{V}_{;\, (a_3 a_4 c)} \right]_\text{3\s perms} \,.
\label{eqn:Amp4AJM}
\end{align}
where $R_{abcd}$ denotes the Riemann curvature tensor derived from the metric $g_{ab}$ in the standard way. We see from these examples that the field space geometry does not address the kinematic factors in the amplitudes. It only provides a geometric interpretation for the coefficients of each kinematic combination that can appear.\footnote{Note that \cref{eqn:Amp3AJM,eqn:Amp4AJM} assume w.l.o.g.\ that $\overline{g}_{ab}$ and $\overline{V}_{;ab}=m_a^2\s \overline{g}_{ab}$ are diagonal. See \cite{Cheung:2021yog,Alonso:2022ffe} for details of how to avoid this assumption with the use of vielbeins.}

\subsubsection{Embedding the Field Space Manifold into the Functional Manifold}
\label{subsubsec:TwoManifolds}

The field space geometry is constructed on the manifold of the field space, while the functional manifold discussed in \cref{subsec:FunctionalManifold} consists of the field configuration space. Therefore, the finite dimensional field space manifold could be identified with a submanifold of the infinite dimensional functional manifold, defined by the restriction that it only contains the constant field configurations.

However, this is not to say that the field space geometry only handles constant field configurations. It addresses arbitrary field configurations by invoking the field maps $\phi^a(x)$ from the spacetime manifold to the field space manifold, inducing a factorized structure of the connection (\cf\ \cref{eqn:FieldSpaceConnection}):
\begin{equation}
\dd\phi^b = \dd x^\mu \left(\partial_\mu\phi^b\right)
\qquad\Longrightarrow\qquad
\dd \left( \frac{\partial}{\partial\phi^c} \right) = \dd x^\mu \Big[ \left(\partial_\mu\phi^b\right) \Gamma_{bc}^a \Big] \left( \frac{\partial}{\partial\phi^a} \right) \,.
\label{eqn:FactorizedConnection}
\end{equation}
The term in the squared bracket can be viewed as a connection that defines a covariant derivative $\mathscr{D}_\mu$ on the spacetime manifold; see \eg~\cite{Alonso:2016oah}. For example, the first derivative of the potential $V_{,\,a}=V_{;\,a}$ is a $(0,1)$-tensor on the field space manifold. Its spacetime covariant derivative is then given by
\begin{equation}
\mathscr{D}_\mu V_{;\,a} = \partial_\mu V_{;\,a} - \Gamma_{ba}^c \left( \partial_\mu \phi^b \right) V_{;\,c}
= \left( \partial_\mu \phi^b \right) \nabla_b V_{;\,a} = \left( \partial_\mu \phi^b \right) V_{;\,ab} \,.
\end{equation}

On the other hand, the functional manifold is formed by all the field maps $\{\phi^a(x)\}$. The bases of its tangent and cotangent space are ``promoted'' from the field space manifold version into (\cf\ \cref{eqn:BasesTrans})
\begin{subequations}
\begin{align}
\frac{\partial}{\partial\phi^a}
\quad\longrightarrow\quad
\frac{\delta}{\delta\phi^a(x)} \,, \\[5pt]
\dd\phi^a
\quad\longrightarrow\quad
\delta\phi^a(x) \,.
\end{align}
\end{subequations}

\subsubsection{Reproducing the Connection on the Field Space Manifold}
\label{subsubsec:Connection}

We would like to reproduce geometric quantities on the field space manifold from quantities on the functional manifold.  To this end, we should restrict the functional manifold quantities onto the submanifold formed by constant field configurations, namely by taking
\begin{equation}
\partial_\mu \phi^a = 0 \,.
\end{equation}
In what follows, we will show how to reproduce the field space manifold connection $\Gamma_{bc}^a$ from $\Chris_{x_1 x_2}^y$, even though the latter does not serve as a connection on the functional manifold. More specifically, we will take the definition of $\Chris_{x_1 x_2}^y$ in \cref{eqn:Chrisdef} and apply it to the theory given by the Lagrangian in \cref{eqn:LagAJM} at the tree level. We then restrict the resulting expression onto the submanifold formed by constant field configurations, and show that this gives us $\Gamma_{bc}^a$.

We begin with the 1PI effective action at tree level, which is just the action:
\begin{equation}
\Gamma [\phi] = S [\phi] = \int \dd^4x 
\left[ - V (\phi) + \frac12\, g_{ab}(\phi) \left( \partial_\mu \phi^a \right) \left( \partial^\mu \phi^b \right) \right]_x \,.
\end{equation}
Here everything in the squared bracket is evaluated at the spacetime point $x$, as indicated by the subscript $x$ shorthand. Note that without following a comma or semicolon, this subscript $x$ is not denoting a functional derivative, but simply denotes evaluating the function at $x$, as in the cases of $\phi^x$ and $J_x$. Note that we are keeping the flavor indices explicit. We need its first functional derivative
\begin{align}
\frac{\delta\Gamma}{\delta\phi^a(x_1)} = - \Big[ g_{ai} \left(\partial^2\phi^i\right) + \left( g_{ai,j} - \tfrac12 g_{ij,a} \right) (\partial_\mu\phi^i) (\partial^\mu\phi^j) + V_{,a} \Big]_{x_1} \,,
\end{align}
its second functional derivative
\begin{align}
\frac{\delta^2\Gamma}{\delta\phi^a(x_1) \delta\phi^b(x_2)} &= - \Big\{
(g_{ab})_{x_1} \big[\partial^2\delta^4(x_1-x_2)\big]
+ \left( g_{ai,b} \partial^2 \phi^i \right)_{x_1} \delta^4(x_1-x_2)
\notag\\[5pt]
&\hspace{20pt}
+ \big[ (g_{ab,i}-g_{ib,a}+g_{ai,b}) (\partial_\mu\phi^i) \big]_{x_1} \big[\partial^\mu\delta^4(x_1-x_2)\big]
\notag\\[5pt]
&\hspace{20pt}
+ \left[ \left( g_{ai,jb} - \tfrac12 g_{ij,ab} \right) (\partial_\mu\phi^i) (\partial^\mu\phi^j) \right]_{x_1} \delta^4(x_1-x_2)
\notag\\[5pt]
&\hspace{20pt}
+ (V_{,ab})_{x_1} \delta^4(x_1-x_2) \Big\} \,,
\end{align}
and its third functional derivative
\begin{align}
\frac{\delta^3\Gamma}{\delta\phi^a(x_1) \delta\phi^b(x_2) \delta\phi^k(z)} &= - \Big\{
(g_{ab,k})_{x_1} \big[\partial^2\delta^4(x_1-x_2)\big] \delta^4(x_1-z)
\notag\\[5pt]
&\hspace{-80pt}
+ (g_{ak,b})_{x_1} \delta^4(x_1-x_2) \big[\partial^2\delta^4(x_1-z)\big]
\notag\\[5pt]
&\hspace{-80pt}
+ (g_{ab,k}-g_{kb,a}+g_{ak,b})_{x_1} \big[\partial_\mu\delta^4(x_1-x_2)\big] \big[\partial^\mu\delta^4(x_1-z)\big]
\notag\\[5pt]
&\hspace{-80pt}
+ \big[ (g_{ab,ik}-g_{ib,ak}+g_{ai,bk}) (\partial_\mu\phi^i) \big]_{x_1} \big[\partial^\mu\delta^4(x_1-x_2)\big] \delta^4(x_1-z)
\notag\\[5pt]
&\hspace{-80pt}
+ \big[ (g_{ai,kb}-g_{ik,ab}+g_{ak,ib}) (\partial_\mu\phi^i) \big]_{x_1} \delta^4(x_1-x_2) \big[\partial^\mu\delta^4(x_1-z)\big]
\notag\\[5pt]
&\hspace{-80pt}
+ \left[ g_{ai,bk} (\partial^2 \phi_i) + \left( g_{ai,jbk} - \tfrac12 g_{ij,abk} \right) (\partial_\mu\phi^i) (\partial^\mu\phi^j) \right]_{x_1} \delta^4(x_1-x_2)\delta^4(x_1-z)
\notag\\[5pt]
&\hspace{-80pt}
+ (V_{,abk})_{x_1} \delta^4(x_1-x_2)\delta^4(x_1-z) \Big\} \,.
\end{align}
Now using the definition in \cref{eqn:Chrisdef} and restricting to the constant field configurations, we get
\begin{align}
\Chris_{ab}^c(x_1, x_2; y) \big|_{\partial_\mu\phi^a=0} &\equiv -\frac{\delta^3\Gamma}{\delta\phi^a(x_1) \delta\phi^b(x_2) \delta\phi^k(z)}\, iD^{kc}(z, y) \Bigg|_{\partial_\mu\phi^a=0} \notag\\[5pt]
&\hspace{-80pt} = \int \dd^4z \bigg\{
g_{ab,k} \big[\partial^2\delta^4(x_1-x_2)\big] \delta^4(x_1-z)
+ g_{ak,b}\s \delta^4(x_1-x_2) \big[\partial^2\delta^4(x_1-z)\big]
\notag\\[5pt]
&\hspace{-50pt}
+ (g_{ab,k}-g_{kb,a}+g_{ak,b}) \big[\partial_\mu\delta^4(x_1-x_2)\big] \big[\partial^\mu\delta^4(x_1-z)\big]
\notag\\[5pt]
&\hspace{-50pt}
+ V_{,abk}\s \delta^4(x_1-x_2)\delta^4(x_1-z) \bigg\}
\int \frac{\dd^4p}{(2\pi)^4}\, e^{-ip(z-y)} \frac{-1}{g_{kc}\s p^2-V_{,kc}} \,.
\end{align}
It is more convenient to take a Fourier transform
\begin{align}
&\int \dd^4x_1\, \dd^4x_2\, \dd^4y\, e^{ip_1x_1+ip_2x_2} e^{-iqy} \left[ G_{ab}^c\left(x_1, x_2; y\right)\big|_{\partial_\mu\phi^a=0} \right]
\notag\\[8pt]
&= (2\pi)^4\delta^4(p_1+p_2-q) \frac{1}{g_{kc}\s q^2-V_{,kc}}
\Big[ \tfrac12 (g_{ka,b}+g_{kb,a}-g_{ab,k})\, q^2 + \tfrac12 (g_{ab,k}-g_{kb,a}+g_{ka,b})\, p_1^2
\notag\\[5pt]
&\hspace{100pt} + \tfrac12 (g_{ab,k}+g_{kb,a}-g_{ka,b})\, p_2^2 - V_{,abk} \Big] \,.
\end{align}
We see that when the potential is absent in the theory, and the external momenta $p_1, p_2$ are on shell $\pon_1^2=\pon_2^2=0$, we indeed reproduce the field space manifold connection:
\begin{align}
&\int \dd^4x_1\, \dd^4x_2\, \dd^4y\, e^{i\pon_1x_1+i\pon_2x_2} e^{-iqy} \left[ G_{ab}^c\left(x_1, x_2; y\right)\big|_{\partial_\mu\phi^a=0} \right] \notag\\[5pt]
&\hspace{80pt}
= (2\pi)^4\delta^4(\pon_1+\pon_2-q)\, \Gamma_{ab}^c \,,
\label{eqn:ChrisReduction}
\end{align}
or equivalently written with the external wavefunctions (\cref{eqn:psixScalar}) as
\begin{equation}
\psi^{x_1}(\pon_1)\, \psi^{x_2}(\pon_2) \left[ G_{ab}^c\left(x_1, x_2; y\right)\big|_{\partial_\mu\phi^a=0} \right]
= R_\eta^{1/2}\, \psi^y(\pon_1+\pon_2)\, \Gamma_{ab}^c \,.
\label{eqn:ChrisReductionpsi}
\end{equation}
When the potential is present, $\Gamma_{ab}^c$ is reproduced from $G_{ab}^c\left(x_1, x_2; y\right)$ in the kinematic limit of large $q^2$. These results demonstrate that $\Chris_{x_1 x_2}^y$ serves as a generalization of $\Gamma_{bc}^a$, even though it does not have a geometric meaning on the functional manifold.

\subsubsection{Reproducing the Geometric Soft Theorem}
\label{subsubsec:SoftTheorem}

A nice result obtained from the field space geometry picture is the so-called geometric soft theorem \cite{Cheung:2021yog}. When applied to the scalar field theory in \cref{eqn:LagAJM} with only the two-derivative term,\footnote{We focus on the zero potential case here for simplicity of the presentation. When the potential term is turned on, the geometric soft theorem is slightly more complicated; see Eq.~(17) in \cite{Cheung:2021yog}. It can be also reproduced in a similar way.} it states that in the soft kinematic limit of the $(n+1)^\text{th}$ leg (labeled by the flavor index $b$ below), the on-shell amplitudes satisfy the following recursion relation
\begin{equation}
\lim_{\qon\to 0}\, \Amp_{a_1 \cdots a_n b} \left( \pon_1, \cdots, \pon_n, \qon \right) = R_\eta^{1/2}\, \nabla_b\, \Amp_{a_1 \cdots a_n} \left( \pon_1, \cdots, \pon_n \right) \,,
\label{eqn:RecursionFieldSpace}
\end{equation}
where $\nabla_b$ is the covariant derivative on the field space manifold; see \cref{eqn:nabladef} for explicit expression. In this subsection, we show that the tensor-like recursion relation in \cref{eqn:RecursionGeo} serves as a generalized version of \cref{eqn:RecursionFieldSpace}, in the sense that it reproduces \cref{eqn:RecursionFieldSpace} when restricted to the submanifold of constant field configurations.\footnote{Note that the residue factor $R_\eta^{1/2}$ in \cref{eqn:RecursionFieldSpace} can be extracted from the analogous all-order expression in \cite{Cheung:2021yog} by changing from a mass basis to a flavor basis index. We assume no mass mixing between flavor eigenstates.}

We begin with the functional derivative part of \cref{eqn:RecursionGeo}. When restricted to the submanifold of constant field configurations, and taking the $q\to 0$ limit, we have
\begin{equation}
\lim_{q\to 0}\; \psi^y(q)\, \tfrac{\delta}{\delta\phi^b(y)}\, \Big[ \Mcal_{a_1 \cdots a_n} ( x_1, \cdots, x_n ) \big|_{\partial_\mu\phi^a=0} \Big]
= R_\eta^{1/2}\, \tfrac{\partial}{\partial\phi^b}\, \Mcal_{a_1 \cdots a_n} \left( x_1, \cdots, x_n \right) \,.
\label{eqn:Der0}
\end{equation}
Now using \cref{eqn:Ampdef}, we get
\begin{align}
\lim_{q\to 0}&\; \Big[ (2\pi)^4 \delta^4(p_1 + \cdots + p_n + q)\, i\Amp_{a_1 \cdots a_n b} \left( p_1, \cdots, p_n, q \right) \Big]
\notag\\[5pt]
&\qquad
= \big[ \psi^{x_1}(p_1) \cdots \psi^{x_n}(p_n)\, \psi^y(q) \big]\, 
\Bigl[ -i\Mcal_{a_1 \cdots a_n b} \left( x_1, \cdots, x_n, y \right) \big|_{J=0} \Bigr]
\notag\\[5pt]
&\qquad
\supset \big[ \psi^{x_1}(p_1) \cdots \psi^{x_n}(p_n)\, \psi^y(q) \big]\, 
\left[ -i \frac{\delta}{\delta\phi^b(y)}\, \Mcal_{a_1 \cdots a_n} \left( x_1, \cdots, x_n \right) \right] \bigg|_{J=0}
\notag\\[5pt]
&\qquad
= R_\eta^{1/2}\, \big[ \psi^{x_1}(p_1) \cdots \psi^{x_n}(p_n) \big]
\left[ -i \frac{\partial}{\partial\phi^b}\, \Mcal_{a_1 \cdots a_n} \left( x_1, \cdots, x_n \right) \right] \bigg|_{J=0}
\notag\\[5pt]
&\qquad
= (2\pi)^4 \delta^4(p_1 + \cdots + p_n)\, R_\eta^{1/2} \frac{\partial}{\partial\phi^b}\, i\Amp_{a_1 \cdots a_n} \left( p_1, \cdots, p_n \right) \,,
\label{eqn:Der1}
\end{align}
or simply
\begin{equation}
\lim_{q\to 0}\; \Amp_{a_1 \cdots a_n b} \left( p_1, \cdots, p_n, q \right)
\supset R_\eta^{1/2}\, \frac{\partial}{\partial\phi^b}\, \Amp_{a_1 \cdots a_n} \left( p_1, \cdots, p_n \right) \,.
\label{eqn:Der2}
\end{equation}
Next let us work out the connection part of \cref{eqn:RecursionGeo}. Taking the momenta to be on-shell, \ie, $p_i = \pon_i$ and $q=\qon$, we can make use of \cref{eqn:ChrisReductionpsi} to get
\begin{align}
& (2\pi)^4 \delta^4(\pon_1 + \cdots + \pon_n + \qon)\, i\Amp_{a_1 \cdots a_n b} \left( \pon_1, \cdots, \pon_n, \qon \right)
\notag\\[5pt]
&\qquad
\supset \big[ \psi^{x_1}(\pon_1) \cdots \psi^{x_n}(\pon_n)\, \psi^y(\qon) \big]
\notag\\[5pt]
&\hspace{80pt}
\times \int \dd^4z \Big[ - \Chris_{b a_1}^c \left( y, x_1; z \right) \big|_{\partial_\mu\phi^a=0} \Big]
\Bigl[ - i\Mcal_{c a_2 \cdots a_n} \left( z, x_2, \cdots, x_n \right) \big|_{J=0} \Bigr]
\notag\\[5pt]
&\qquad
= R_\eta^{1/2}\, \big[ \psi^z(\pon_1+\qon)\, \psi^{x_2}(\pon_2) \cdots \psi^{x_n}(\pon_n) \big]\, \Gamma_{b a_1}^c
\Bigl[ i\Mcal_{c a_2 \cdots a_n} \left( z, x_2, \cdots, x_n \right) \big|_{J=0} \Bigr]
\notag\\[5pt]
&\qquad
= (2\pi)^4 \delta^4(\pon_1 + \cdots + \pon_n + \qon)\, R_\eta^{1/2} \left( -\Gamma_{b a_1}^c \right)
i\Amp_{c a_2 \cdots a_n} \left( \pon_1 + \qon, \pon_2, \cdots, \pon_n \right) \,.
\label{eqn:Chris1}
\end{align}
Taking the soft limit, this reads
\begin{equation}
\lim_{\qon\to 0}\; \Amp_{a_1 \cdots a_n b} \left( \pon_1, \cdots, \pon_n, \qon \right)
\supset R_\eta^{1/2} \left( -\Gamma_{b a_1}^c \right) \Amp_{c a_2 \cdots a_n} \left( \pon_1, \pon_2, \cdots, \pon_n \right) \,.
\label{eqn:Chris2}
\end{equation}
Combining \cref{eqn:Der2,eqn:Chris2}, we obtain \cref{eqn:RecursionFieldSpace}.

\subsubsection{Revisiting Vanishing Curvature for Functional Geometry}
\label{sec:VanishingCurvatureFieldSpaceGeo}

We can gain some insight into why we are finding that functional geometry has zero curvature (see \cref{subsubsec:ZeroCurvature}) by comparing with the case of field space geometry. Consider the expression in \cref{eqn:Amp4AJM} for the four-point amplitude written using field space geometry. The amplitude is written as a sum of several terms, and each can be expressed as a geometric quantity multiplied by kinematic dependence. Under a non-derivative field redefinition, each term here is individually invariant. On the other hand, when a derivative field redefinition is carried out, each term alone will no longer have a well-defined geometric meaning. However, the total amplitude is of course still invariant. A repackaged expression of \cref{eqn:Amp4AJM} is desired to make this invariance manifest, which would serve as a generalization of the field space geometry. This is what we hoped (and failed) to accomplish by introducing functional geometry.

Taking a closer look at the expression in \cref{eqn:Amp4AJM}, we note that it contains two types of geometric quantities: some of its terms are fully determined by the Riemann curvature tensor, which is the intrinsic geometry of the field space manifold endowed with the metric $g_{ab}(\phi)$, while others depend on external input functions on the manifold, such as the potential $V(\phi)$. From this point of view, a generalization of the field space geometry will repackage \cref{eqn:Amp4AJM} still into these two types of geometric quantities, under the new notion of geometry. Apparently, what the functional geometry picture has done is to package everything into the second type.  The intrinsic geometry is trivialized since the curvature vanishes, and the amplitude is fully determined by an external input function, namely the 1PI effective action $\Gamma[\phi]$.

\subsection{Exploring Modified Source Terms}
\label{subsec:Explorations}

In this section we consider if modifications of the source term that appears in the path integral can change the conclusions about the lack of curvature for functional geometry. Modifications of the source term in the partition function change the off-shell behavior of correlators but leave amplitudes invariant. This statement is a key feature of the traditional argument for field redefinition invariance of amplitudes, see \eg\ \cite{Manohar:2018aog} and \cref{appsec:Invariance}. The freedom to modify the source term has also been used by Vilkovisky \cite{Vilkovisky:1984st} and DeWitt \cite{DeWitt:1984sjp, DeWitt:1985sg} to define specific effective actions whose correlators transform covariantly with respect to transformations in field space; the same freedom may be useful here for removing evanescent terms in configuration space.

We define a new partition function $\tZ[J]$, which differs from \cref{eqn:ZWdef} by an additional source term $\delta T[\eta, J]$:
\begin{equation}
\tZ[J] \equiv \int\Dcal \eta\, e^{iS + i J_x \eta^x + i \delta T[\eta, J]} \;.
\label{eqn:tZdef}
\end{equation}
We assume that $\delta T[\eta, J]$ has a smooth dependence on $\eta$, which admits the following functional expansion
\begin{equation}
\delta T = \tJ_{y_1}^{(1)} (\tphi-\eta)^{y_1} + \tJ_{y_1y_2}^{(2)} (\tphi-\eta)^{y_1} (\tphi-\eta)^{y_2} + \tJ_{y_1y_2y_3}^{(3)} (\tphi-\eta)^{y_1} (\tphi-\eta)^{y_2} (\tphi-\eta)^{y_3} + \ldots
\label{eqn:Tdef}
\end{equation}
where the coefficients $\tJ^{(i)}$ are $\eta$ independent, but functionals of $J$. Note that $\tphi$ here is a functional of $J$, which is implicitly determined through its definition
\begin{equation}
\tphi^y \equiv \expv{ \eta^{y} }_{\delta T, J} \equiv \frac{\int \Dcal \eta \, \eta^y \, e^{iS + i J_x \eta^x + i \delta T}}{\int \Dcal \eta \, e^{iS + i J_x \eta^x + i \delta T}} \,.
\label{eqn:tphidef}
\end{equation}
It is expected that the $\tJ^{(i)}_{y_1 \dots y_k}$ are local, in that they are only supported when $y_1=y_2=\ldots=y_k$, but the following analysis does not rely on this.

Therefore, \cref{eqn:Tdef} is an expansion of the $\eta$ dependence about its quantum vev. This is done to reduce the size of the ensuing expressions, and we can make this shift without loss of generality. \cref{eqn:Tdef} could be rewritten as an expansion about $\eta=0$: as $\tphi$ is a functional of $J$, the $\tphi$ terms can be absorbed in \cref{eqn:Tdef} through redefinitions of the $\tJ^{(i)}$, up to an $\eta$-independent phase which drops out of all correlators.\footnote{Note that the definition of $\tphi^y$, \cref{eqn:tphidef}, is self-referential, but it can be iteratively solved to an arbitrarily high power of $J$ and $\tJ^{(i)}$.}

We define a set of analogous tilded quantities that are modified with respect to the quantities in previous sections due to the presence of the extra source terms:
\begin{subequations}\label{eqn:modifiedQuantities}
\begin{align}
\tD^{xy} &\equiv \expv{ \eta^x \eta^y }_{\delta T,J,\,\text{conn}} \,, \\[5pt]
-i\tcM_{x_1 \cdots x_n} &\equiv \bigg( \prod_i \tD^{-1}_{x_i y_i} \bigg)
\expv{ \eta^{y_1} \cdots \eta^{y_n} }_{\delta T,J,\,\text{conn}} \,, \\[5pt]
\tG^z_{x_1x_2} &\equiv i \tD^{zy}\, \tcM_{y x_1 x_2} \,, \\[5pt]
\tnabla_y\, \tcM_{x_1 \cdots x_n} &\equiv \frac{\delta}{\delta \tphi^y}\, \tcM_{x_1 \cdots x_n}
- \sum_{i=1}^n \tG^{z}_{y x_i}\, \tcM_{x_1 \cdots \hat{x}_i z \cdots x_n}  \,.
\end{align}
\end{subequations}
Working to first order in the extra source terms, any modified correlator can be expanded in terms of unmodified ones
\begin{align}
\expv{ (\cdots) }_{\delta T,J} &\equiv \frac{\int \Dcal \eta\, (\cdots)\, e^{iS + iJ_x\eta^x + i\delta T}}{\int \Dcal\eta\, e^{iS + iJ_x\eta^x + i\delta T}} \notag\\[5pt]
&= \frac{\int \Dcal\eta\, (\cdots)  (1 + i\delta T)\, e^{iS + iJ_x\eta^x}}{\int\Dcal\eta\, (1 + i\delta T)\, e^{iS + iJ_x\eta^x}} + \ord\big(\delta T^2\big)
\notag\\[8pt]
&= \expv{ (\cdots) }_{J} + \expv{ (\cdots)  i \delta T}_{J}
- \expv{ (\cdots) }_{J} \expv{ i \delta T }_{J} + \ord\big(\delta T^2\big) \,.
\end{align}
By further expanding $\tcM$, $\tG$, and $\tnabla \tcM$ (which depend on products of correlators), the linear dependence in $\delta T$ of the quantities in \cref{eqn:modifiedQuantities} then follows.
We test the resulting modifications to the recursion relation by computing the difference
\begin{equation}
\mathfrak{D} = -i \left( \tnabla_{x_4}\, \tcM_{x_1 x_2 x_3} - \tcM_{x_1 x_2 x_3 x_4} \right) \,, 
\end{equation}
which is zero in the unmodified path integral. This gives
\newcommand{\hl}[1]{\underline{#1}}
\begin{align}
\mathfrak{D} &= \left\{ \tJ^{(2)}_{y_1y_2,x_4} + 3 \tJ^{(3)}_{y_1y_2x_4} \right\} D^{y_1d_1} D^{y_2d_2} \Big(  \hl{i \cM_{d_1d_2f_1} D^{f_1f_2} \cM_{f_2x_1x_2x_3}} + \hl{\cM_{d_1d_2x_1x_2x_3}}
\notag\\[5pt]
&\qquad
- \left[ \cM_{d_1d_2f_1} D^{f_1f_2} \cM_{f_2f_3x_1} D^{f_3f_4} \cM_{f_4x_2x_3} -i \cM_{d_1 d_2 f_1 x_1} D^{f_1f_2} \cM_{f_2 x_2 x_3} + \text{cycs}\right] \Big)
\notag\\[5pt]
&\quad
- \left\{ \tJ^{(3)}_{y_1y_2y_3,x_4} + 4 \tJ^{(4)}_{y_1y_2y_3x_4} \right\} D^{y_1d_1} D^{y_2d_2} D^{y_3d_3} \notag\\[5pt]
&\qquad \times \Big(  \hl{i \cM_{d_1d_2d_3f_1} D^{f_1f_2} \cM_{f_2x_1x_2x_3}} + \hl{\cM_{d_1d_2d_3x_1x_2x_3}} \notag\\[5pt]
&\qquad \quad
- \left[\cM_{d_1d_2d_3f_1} D^{f_1f_2} \cM_{f_2f_3x_3} D^{f_3f_4} \cM_{f_4x_1x_2}  - i \cM_{d_1 d_2 d_3 f_1 x_1} D^{f_1f_2} \cM_{f_2 x_2 x_3} + \text{cycs}\right] \Big)
\notag\\[5pt]
&\quad
- \Big[ 3 \left\{ \tJ^{(3)}_{y_1y_2x_1,x_4}  + 4 \tJ^{(4)}_{y_1y_2x_1x_4} \right\} D^{y_1d_1} D^{y_2d_2} \notag\\[5pt] 
&\qquad \times \left( \hl{\cM_{d_1d_2x_2x_3}} + i \cM_{d_1d_2f_1} D^{f_1f_2} \cM_{f_2x_2x_3} \right) + \text{cycs}\Big]
- \hl{6 \left\{ \tJ^{(3)}_{x_1x_2x_3,x_4} + 4 \tJ^{(4)}_{x_1x_2x_3x_4} \right\}}
\notag\\[5pt]
&\quad
+ \text{[terms $\propto$ derivatives of $\tJ^{(4)}$]} + \text{[terms $\propto$ $\tJ^{(i)}$ for $i>4$]}+ \ord \big(\delta T^2\big) \,,
\label{eqn:ModifiedRR}
\end{align}
where `cycs' refers to the terms generated by cyclically permuting $x_1,x_2,x_3$, and an index after a comma denotes a functional derivative with respect to $\tphi$, for example $\tJ^{(2)}_{y_1y_2,x_4} \equiv \frac{\delta}{\delta \tphi^{x_4}}\tJ^{(2)}_{y_1y_2}$. (The underlining in this expression has no mathematical meaning and will be used simply to identify terms in the discussion below.)

We note that the non-zero right-hand side of \cref{eqn:ModifiedRR} cannot be wholly absorbed by a redefinition of the ``connection'' $\tG^{z}_{x_1x_2} \to \tG^{z}_{x_1x_2} + \delta\tG^{z}_{x_1x_4}$, for some $\delta\tG^{z}_{x_1x_2}$ linear in $\delta T$. This redefinition would exclusively generate terms of the form $-\delta\tG^{z}_{x_1x_4} \cM_{z x_2 x_3} + \text{cycs}$. However, the underlined terms in \cref{eqn:ModifiedRR} do not contain a piece $\cM_{z x_2 x_3}$ for some dummy index $z$, so they could not be set to zero by such a redefinition.

Nonetheless, the parts of $\mathfrak{D}$ shown in \cref{eqn:ModifiedRR} can be set to zero for $\tJ^{(i)}$ satisfying the conditions
\begin{subequations}
\begin{align}
\tJ^{(2)}_{y_1y_2,y_3} + 3 \tJ^{(3)}_{y_1y_2y_3} &= \ord\big(\delta T^2\big) \,, \\[5pt]
\tJ^{(3)}_{y_1y_2y_3,y_4} + 4 \tJ^{(4)}_{y_1y_2y_3y_4} &= \ord\big(\delta T^2\big) \,.
\end{align}
\end{subequations}
These describe some necessary conditions that additional source terms must satisfy to maintain the recursion relation between correlators. These modifications, in analog with the Vilkovisky and DeWitt effective actions, have the potential to change the transformations of correlators under field redefinitions. This leaves open the exciting possibility that a judicious choice of $\tJ^{(i)}$ can remove the evanescent terms in $G$ and $\cM$, which prevent a clear geometric interpretation of this formalism. We leave this for future work.

%%%%%%%%%%%%%%%%%%%%%%%%%%%%%%%%%%%%%%%%%%%%%%%%%%%%%%%%%%%%%%%%%%%%%%%%%%%%%%%%
\section{Conclusions and Outlook}
\label{sec:Conclusions}
%%%%%%%%%%%%%%%%%%%%%%%%%%%%%%%%%%%%%%%%%%%%%%%%%%%%%%%%%%%%%%%%%%%%%%%%%%%%%%%%

In this paper, we provided a new perspective on the covariance properties of generalized amplitudes under field redefinitions. We proved a result that connects the transformation properties of the 1PI effective action to the transformation of the generalized amplitudes, which we called the \emph{transformation lemma}. We then showed that this result can be applied to demonstrate the invariance of on-shell amplitudes under field redefinitions for scalar field theories up to one-loop order.

The covariance properties of these objects is highly suggestive of an underlying geometric interpretation, that we refer to as functional geometry. We explored the ways in which this functional geometry construction succeeds and where it does not. In particular, the curvature invariants (as computed by naively following the strategy for Riemannian geometry) vanish, and it is currently unclear if a modified approach (for example, adjusting the source term in the path integral) can resolve this issue. Nonetheless, we showed that the functional geometry does reduce to the field space geometry in the appropriate limits, which provides some evidence that this approach is on the right track. See also \cite{Kluth:2023sey} for recent progress on using n-particle irreducible effective actions to study a geometric interpretation.

Generally speaking, any improved understanding of field redefinition freedom in quantum field theory improves our understanding of the physical content of its Lagrangian. It also provides insight into the intricate structure of its amplitudes, which project out these field redefinition redundancies in a non-trivial way. There are many open questions that we would like to explore in the future.

We expect that the condition on the transformation of the effective action, \cref{eqn:GammaRelation}, should hold to all orders in perturbation theory. Since we have only shown this up to one loop, it would be very interesting to understand how this holds at two loops (and beyond), which could help lead to an all-orders result. It is possible that a looser assumption than \cref{eqn:GammaRelation} would still result in the transformation of the correlator given in \cref{eqn:McalRelation}; understanding the minimal possible conditions on the transformation of the effective action could further constrain the edge cases of the allowed space of field redefinitions. It would also be useful to extend our results explicitly to fermionic theories, as well as to understand how the gauge redundancy in gauge theories (whose behavior is in many ways similar to the field redefinition freedom in an ungauged theory) can be included in our framework.

It is worth considering the assumptions we have imposed on the possible space of field redefinitions. In principle, the transformation lemma, \cref{eqn:McalRelation}, holds for any invertible, infinitely differentiable (\ie\ smooth) functional $\phi[\tphi]$. However, in order for the ``evanescent term'' $U$ to be projected out in the amplitude by LSZ reduction, we have assumed many properties in our treatment of the on-shell states. In particular, in \cref{subsec:Amplitudes}, we assume properties of the pole structure of two-point correlator (whence the usual restriction that field redefinitions should be local, in order not to disturb said pole structure), as well as Poincar\'e invariance. This latter requirement of preserving spacetime symmetry is unnecessary, and there are many examples of non-Poincar\'e invariant field theories that have a well-defined $S$-matrix. It would be interesting to extend the results of this paper to such non-Poincar\'e invariant theories.

One interesting intermediate step would be to define the wavefunctions $\psi$ for $J \neq 0$, and therefore define a $J$-dependent amplitude via the LSZ reduction formula \cref{eqn:Ampdef}, which would describe scattering about an arbitrary spatially-dependent background. Understanding the background dependence could be useful for investigating various IR constraints on EFTs. Knowing the $J$-dependence of the amplitude would also allow us to write a functional recursion relation for the amplitudes themselves, \ie, $\Amp$ in addition to $\Mcal$, which could serve as a generalization of the expressions in \cite{Cheung:2021yog} away from the soft and spatially constant background limit.

Finally, the true nature of the functional geometry remains to be discovered. Perhaps this can be accomplished by finding the appropriate source term in the path integral as explored above. Another approach would be to find a way to quotient out the evanescent terms in order to construct the functional manifold directly. It would also be fascinating to understand if there is a connection between functional geometry and recent progress understanding EFTs in terms of Lagrange space \cite{Craig:2023wni} and/or jet bundles~\cite{Craig:2023hhp, Alminawi:2023qtf}. Clearly, we have only begun to address some of the most fundamental questions regarding the connections between EFTs and geometry.

\acknowledgments

We thank Nathaniel Craig for collaboration during the early stages of this work.
We would like to thank Andreas Helset and Aneesh Manohar for useful conversations.
T.~Cohen is supported by the U.S.~Department of Energy under grant number DE-SC0011640.
%+
X.~Lu is supported by the U.S.~Department of Energy under grant number DE-SC0009919.
D.~Sutherland acknowledges support from the Institute for Particle Physics Phenomenology Associateship Scheme.

\appendix
\section*{Appendix}

\section{Amplitude Invariance From the Path Integral}
\label{appsec:Invariance}

In this appendix, we briefly review the argument for amplitude invariance under field redefinitions from the path integral point of view. This is largely repeating Section 6.2 in \cite{Manohar:2018aog}. We include this appendix to make this paper self-contained.

To compute the amplitudes for a theory given by $S[\eta]$, one can start with the generating functional $W[J]$ defined in \cref{eqn:ZWdef}:
\begin{equation}
e^{iW[J]} \equiv \int \Dcal\s\eta\, \exp \left\{ iS[\eta] + i\int \dd^4x\, J(x)\s \eta(x) \right\} \,,
\end{equation}
which generates the connected correlation functions. Making an integration variable change
\begin{equation}
\eta = f \big[ \tilde\eta \big] \,,
\end{equation}
we get the same quantity rewritten as
\begin{align}
e^{iW[J]} &= \int \det\left( \frac{\delta\eta}{\delta\tilde\eta} \right) \Dcal\s\tilde\eta\, \exp \left\{
i S \Big[ f \big[ \tilde\eta \big] \Big] + i\int \dd^4x\, J(x)\s f \big[ \tilde\eta \big](x) \right\}
\notag\\[5pt]
&= \int \Dcal\s\tilde\eta\, \exp \left\{
i \left( S \Big[ f \big[ \tilde\eta \big] \Big] - i\log\det\left( \frac{\delta\eta}{\delta\tilde\eta} \right) \right) + i\int \dd^4x\, J(x)\s f \big[ \tilde\eta \big](x) \right\} \,.
\end{align}
Now, consider a slightly different generating functional $W_1[J]$:
\begin{equation}
e^{iW_1[J]} = \int \Dcal\s\tilde\eta\, \exp \left\{
i \left( S \Big[ f \big[ \tilde\eta \big] \Big] - i\log\det\left( \frac{\delta\eta}{\delta\tilde\eta} \right) \right) + i\int \dd^4x\, J(x)\s \tilde\eta(x) \right\} \,,
\label{eqn:W1def}
\end{equation}
where the difference is due to the last term in the exponent. As $W_1[J] \ne W[J]$, it generates a set of connected correlation functions that are different from the original theory $S[\eta]$. However, the only difference between $W_1[J]$ and $W[J]$ is how the source field $J(x)$ is coupled to the theory:
\begin{equation}
\int \dd^4x\, J(x)\, f \big[ \tilde\eta \big] (x) 
\qquad\text{versus}\qquad
\int \dd^4x\, J(x)\, \tilde\eta (x) \,.
\end{equation}
In such cases, for legitimate field redefinitions $f\big[ \tilde\eta \big]$, it is understood \cite{Arzt:1993gz, Manohar:2018aog} that upon the LSZ reduction procedure, they yield the same on-shell amplitudes. Therefore, we see from \cref{eqn:W1def} that for the purposes of computing the on-shell amplitudes for the theory $S[\eta]$, one can alternatively work with a new theory given by the action $\tilde{S}[\tilde\eta]$:
\begin{equation}
\tilde{S}[\tilde\eta] = S \Big[ f \big[ \tilde\eta \big] \Big] - i\log\det\left( \frac{\delta\eta}{\delta\tilde\eta} \right) \,.
\label{eqn:Stilde}
\end{equation}
Note that the second piece from the Jacobian is one-loop sized. For tree-level calculations, one can ignore it and simply use the first term above as the new theory. For loop-level calculations, if one works with dimensional regularization, the second piece above also vanishes due to it being a scaleless integral (except for anomalous fermion chiral transformations); see \eg\ Ref.~\cite{Arzt:1993gz} for more detailed discussions. In this paper, to make our statement independent of the choice of regularization scheme, we keep the second piece above for the loop-level discussions.

\end{spacing}

\begin{spacing}{1.09}
\addcontentsline{toc}{section}{\protect\numberline{}References}%
\bibliographystyle{JHEP}
\bibliography{GeometryEFTs}
\end{spacing}

\end{document}